\documentclass{iopjournal}

\usepackage[utf8]{inputenc}
\usepackage{lmodern}
\usepackage{subcaption}
\usepackage{multirow}
\usepackage{pgfplots}
\pgfplotsset{compat=newest}
\usepackage{tikz}
\usetikzlibrary{patterns}
\usepackage{graphicx}
\usepackage{amssymb}
\usepackage{xspace}
\usepgfplotslibrary{fillbetween}

\usepackage{url}
\usepackage{hyperref}
\usepackage{xcolor}
\usepackage{amsmath}
\usepackage{enumitem}
\usepackage{lineno}
\usepackage{ragged2e}

\begin{document}
\newcommand{\bs}{\boldsymbol}
\newcommand{\nn}{\nonumber}
\newcommand{\ml}{\mathcal}
\newcommand{\ma}{\mathrm}

\newcommand{\dndeta}{\mbox{d$N_{\rm ch}$/d$\eta$}\xspace}

\newcommand{\LRP}{Long Range Plan}
\newcommand{\lhc}{\mbox{LHC}\xspace}
\newcommand{\rhic}{\mbox{RHIC}\xspace}
\newcommand{\sphenix}{\mbox{sPHENIX}\xspace}
\newcommand{\phenix}{\mbox{PHENIX}\xspace}
\newcommand{\phobos}{\mbox{PHOBOS}\xspace}
\newcommand{\brahms}{\mbox{BRAHMS}\xspace}
\newcommand{\alice}{\mbox{ALICE}\xspace}
\newcommand{\cms}{\mbox{CMS}\xspace}
\newcommand{\atlas}{\mbox{ATLAS}\xspace}
\newcommand{\starexp}{\mbox{STAR}\xspace}

\newcommand{\mvtx}{\mbox{MVTX}\xspace}
\newcommand{\intt}{\mbox{INTT}\xspace}
\newcommand{\tpc}{\mbox{TPC}\xspace}
\newcommand{\tpot}{\mbox{TPOT}\xspace}
\newcommand{\emcal}{\mbox{EMCAL}\xspace}
\newcommand{\ihcal}{\mbox{iHCAL}\xspace}
\newcommand{\ohcal}{\mbox{oHCAL}\xspace}
\newcommand{\mbd}{\mbox{MBD}\xspace}
\newcommand{\zdc}{\mbox{ZDC}\xspace}
\newcommand{\sepd}{\mbox{sEPD}\xspace}
\newcommand{\smd}{\mbox{SMD}\xspace}

\newcommand{\minbias}{\textsc{Min. Bias}\xspace}

\newcommand{\hic}{\mbox{$A$$+$$A$}\xspace}
\newcommand{\ee}{\mbox{$e^{+}$$+$$e^{-}$}\xspace}
\newcommand{\ep}{\mbox{$e$$+$$p$}\xspace}
\newcommand{\eA}{\mbox{$e$$+$$A$}\xspace}
\newcommand{\AuAu}{\mbox{Au$+$Au}\xspace}
\newcommand{\uu}{\mbox{U$+$U}\xspace}
\newcommand{\apa}{\mbox{A$+$A}\xspace}
\newcommand{\auau}{\mbox{Au$+$Au}\xspace} 
\newcommand{\alal}{\mbox{Al$+$Al}\xspace} 
\newcommand{\oo}{\mbox{O$+$O}\xspace} 
\newcommand{\agag}{\mbox{Ag$+$Ag}\xspace} 
\newcommand{\cucu}{\mbox{Cu$+$Cu}\xspace} 
\newcommand{\xexe}{\mbox{Xe$+$Xe}\xspace} 
\newcommand{\raa}{\mbox{$R_{\rm AA}$}\xspace}
\newcommand{\pbpb}{\mbox{Pb$+$Pb}\xspace} 
\newcommand{\ppb}{\mbox{p$+$Pb}\xspace} 
\newcommand{\pdau}{\mbox{$p(d)$$+$Au}\xspace} 
\newcommand{\aj}{\mbox{$A_J$}\xspace} 
\newcommand {\pp}{\mbox{$p$$+$$p$}\xspace}
\newcommand {\ppbar}{\mbox{$p$$+$$\overline{p}$}\xspace}
\newcommand{\pT}{\mbox{${p_T}$}\xspace}
\newcommand{\jpsi}{\mbox{$J/\psi$}\xspace}
\newcommand{\ups}{\mbox{$\Upsilon$}\xspace}
\newcommand{\sqrts}{\mbox{$\sqrt{s}$}\xspace}
\newcommand{\sqrtsnn}{\mbox{$\sqrt{s_{\rm NN}}$}\xspace}
\newcommand{\npart}{$N_{\mathrm{part}}$\xspace}
\newcommand{\ncoll}{$N_{\mathrm{coll}}$\xspace}
\newcommand{\qgp}{\mbox{quark-gluon plasma}\xspace}
\newcommand{\jt}{\mbox{$J_T$}} 
\newcommand{\qhat}{\mbox{$\hat{q}$}\xspace}
\newcommand{\Qsqr}{\mbox{$Q^2$}} 
\newcommand{\CuCu}{\mbox{Cu$+$Cu}\xspace}
\newcommand{\PbPb}{\mbox{Pb$+$Pb}\xspace} 
\newcommand{\pPb}{\mbox{$p$$+$Pb}\xspace}
\newcommand{\gjet}{\mbox{$\gamma$-jet}\xspace}
\newcommand{\Qmax}{\mbox{$Q_{\max}$}} 
\newcommand{\ET}{\mbox{$E_T$}}
\newcommand{\Et}{\mbox{$E_T$}} 
\newcommand{\kt}{\mbox{$k_T$}}
\newcommand{\RAA}{\mbox{$R_{\mathrm{AA}}$}\xspace} 
\newcommand{\IAA}{\mbox{$I_{AA}$}}
\newcommand{\pt}{\mbox{${p_{\mathrm{T}}}$}\xspace}
\newcommand{\nb}{\mbox{nb$^{-1}$}\xspace}
\newcommand{\pb}{\mbox{pb$^{-1}$}\xspace}
\newcommand{\fb}{\mbox{fb$^{-1}$}\xspace}

\renewcommand{\AA}{A+A}
\newcommand{\AB}{A+B}
\newcommand{\pA}{p+A}
\newcommand{\dA}{d+A}
\newcommand{\HeA}{$^3$He+A}
\newcommand{\RuRu}{Ru+Ru}
\newcommand{\ZrZr}{Zr+Zr}
\newcommand{\OO}{O+O}
\newcommand{\ArAr}{Ar+Ar}
\newcommand{\UU}{U+U}
\newcommand{\pO}{p+O}
\newcommand{\HeAu}{$^3$He+Au}
\newcommand{\XeXe}{Xe+Xe}
\newcommand{\SiW}{Si+W}
\newcommand{\pAr}{p+Ar}
\newcommand{\PbAr}{Pb+Ar}

\newcommand{\rpa}{$R_{p\rm{A}}$}
\newcommand{\rda}{$R_{d\rm{A}}$}
\newcommand{\rcp}{$R_{\rm CP}$}

\newcommand{\Dzero}{$D^{0}$\xspace}
\newcommand{\dTokpi}{$D^{0}\rightarrow K^-\pi^+$\xspace}
\newcommand{\mkpi}{$m_{K^-\pi^+}$\xspace}
\newcommand {\bbbar}{\mbox{$b\overline{b}$}\xspace}
\newcommand {\ccbar}{\mbox{$c\overline{c}$}\xspace}

\newcommand{\highpt}{high-${\rm p_{_{T}}}$}
\newcommand{\lessim}{{\stackrel{<}{\sim}}} \newcommand{\eqnpt}{p_T}
\newcommand{\dAu}{\mbox{$d$$+$Au}\xspace}
\newcommand{\pAu}{\mbox{$p$$+$Au}\xspace}
\newcommand{\pau}{\mbox{$p$$+$Au}\xspace}
\newcommand{\pAl}{\mbox{$p$$+$Al}\xspace}
\newcommand{\gevsq}{\mbox{${\rm~GeV}^2$}\xspace}
\newcommand{\fastjet}{\mbox{\sc FastJet}\xspace}
\newcommand{\hepmc}{\mbox{\sc HepMC2}\xspace}
\newcommand{\geant}{\mbox{\sc Geant4}\xspace}
\newcommand{\antikt}{\mbox{anti-$k_T$}\xspace}
\newcommand{\pythia}{\mbox{\sc Pythia8}\xspace}
\newcommand{\funforall}{\mbox{\sc Fun4All}\xspace}
\newcommand{\kfparticle}{\mbox{\sc KFParticle}\xspace}
\newcommand{\decayfinder}{\mbox{\sc DecayFinder}\xspace}
\newcommand{\hftrackeff}{\mbox{\sc HFTrackEfficiency}\xspace}
\newcommand{\rapgap}{\mbox{\sc Rapgap}\xspace}
\newcommand{\milou}{\mbox{\sc Milou}\xspace}
\newcommand{\pyquen}{\mbox{\tt Pyquen}\xspace}
\newcommand{\hijing}{\mbox{\tt HIJING}\xspace}
\newcommand{\ampt}{\mbox{\tt AMPT}\xspace}
\newcommand{\epos}{\mbox{\tt EPOS4}\xspace}
\newcommand{\jewel}{\mbox{\tt Jewel}\xspace}
\newcommand{\roofit}{\mbox{\sc RooFit}\xspace}
\newcommand{\roounfold}{\mbox{\sc RooUnfold}\xspace}
\newcommand{\beetle}{\mbox{\sc Beetle}\xspace}
\newcommand{\gj}{\mbox{$\gamma$+jet}\xspace}
\newcommand{\gh}{\mbox{$\gamma$+hadron}\xspace}
\newcommand{\martinimusic}{\mbox{\sc Martini+Music}\xspace}
\newcommand{\martini}{\mbox{\sc Martini}\xspace}
\newcommand{\music}{\mbox{\sc Music}\xspace}
\newcommand{\Ephenix}{Electron-Ion Collider (EIC) detector built
  around the BaBar magnet and sPHENIX calorimetry\xspace}
\newcommand{\ephenix}{EIC detector built around the BaBar magnet and
  sPHENIX calorimetry\xspace} 
\newcommand{\refdesign}{reference design\xspace}
\newcommand{\refconfig}{reference configuration\xspace}
\newcommand{\dijet}{\mbox{dijet}\xspace}
\newcommand{\fake}{\mbox{fake}\xspace}
\newcommand{\fast}{\mbox{fast}\xspace}
\newcommand{\veryfast}{\mbox{very fast}\xspace}
\newcommand{\epem}{\mbox{$e^+e^-$}\xspace}
\newcommand{\onewidth}{0.6\linewidth}
\newcommand{\twowidth}{0.48\linewidth}
\newcommand{\threewidth}{0.32\linewidth}

\newcommand{\egoing}{\mbox{electron-going}\xspace}
\newcommand{\hgoing}{\mbox{hadron-going}\xspace}
\newcommand{\egodir}{electron-going direction\xspace}
\newcommand{\hgodir}{hadron-going direction\xspace}
\newcommand{\bigcell}[2]{\begin{tabular}{@{}#1@{}}#2\end{tabular}}

\def\sPlot{\mbox{\em sPlot}\xspace}
\def\sPlots{\mbox{\em sPlots}\xspace}
\def\sWeights{\mbox{\em sWeights}\xspace}
%%%%%%%%%%%%%%%%%%%%%%%%%%%%%%%%%%%%%%%%%%%%%%%%%%
% Units
%%%%%%%%%%%%%%%%%%%%%%%%%%%%%%%%%%%%%%%%%%%%%%%%%%
%\newcommand{\unit}[1]{\ensuremath{\rm\,#1}\xspace}          % {kg}

%% Energy and momentum
\newcommand{\tev}{\ensuremath{\mathrm{\,Te\kern -0.1em V}}\xspace}
\newcommand{\gev}{\ensuremath{\mathrm{\,Ge\kern -0.1em V}}\xspace}
\newcommand{\mev}{\ensuremath{\mathrm{\,Me\kern -0.1em V}}\xspace}
\newcommand{\kev}{\ensuremath{\mathrm{\,ke\kern -0.1em V}}\xspace}
\newcommand{\ev}{\ensuremath{\mathrm{\,e\kern -0.1em V}}\xspace}
\newcommand{\gevc}{\ensuremath{{\mathrm{\,Ge\kern -0.1em V\!/}c}}\xspace}
\newcommand{\mevc}{\ensuremath{{\mathrm{\,Me\kern -0.1em V\!/}c}}\xspace}
\newcommand{\gevcc}{\ensuremath{{\mathrm{\,Ge\kern -0.1em V\!/}c^2}}\xspace}
\newcommand{\gevgevcccc}{\ensuremath{{\mathrm{\,Ge\kern -0.1em V^2\!/}c^4}}\xspace}
\newcommand{\mevcc}{\ensuremath{{\mathrm{\,Me\kern -0.1em V\!/}c^2}}\xspace}

%% Distance and area
\def\km   {\ensuremath{\rm \,km}\xspace}
\def\m    {\ensuremath{\rm \,m}\xspace}
\def\cm   {\ensuremath{\rm \,cm}\xspace}
\def\cma  {\ensuremath{{\rm \,cm}^2}\xspace}
\def\mm   {\ensuremath{\rm \,mm}\xspace}
\def\mma  {\ensuremath{{\rm \,mm}^2}\xspace}
\def\mum{\ensuremath{\upmu\mathrm{m}}\xspace}
\def\muma {\ensuremath{\rm \,\upmu\mathrm{m}^2}\xspace}
\def\nm   {\ensuremath{\rm \,nm}\xspace}
\def\fm   {\ensuremath{\rm \,fm}\xspace}
\def\barn{\ensuremath{\rm \,b}\xspace}
\def\barnhyph{\ensuremath{\rm -b}\xspace}
\def\mbarn{\ensuremath{\rm \,mb}\xspace}
\def\mub{\rm \,\textmu b\xspace}
\def\mbarnhyph{\ensuremath{\rm -mb}\xspace}
\def\nb {\ensuremath{\rm \,nb}\xspace}
\def\invnb {\ensuremath{\mbox{\,nb}^{-1}}\xspace}
\def\pb {\ensuremath{\rm \,pb}\xspace}
\def\invpb {\ensuremath{\mbox{\,pb}^{-1}}\xspace}
\def\fb   {\ensuremath{\mbox{\,fb}}\xspace}
\def\invfb   {\ensuremath{\mbox{\,fb}^{-1}}\xspace}
\def\khz   {\ensuremath{\mbox{\,kHz}}\xspace}
\def\mhz   {\ensuremath{\mbox{\,MHz}}\xspace}
\def\ghz   {\ensuremath{\mbox{\,GHz}}\xspace}
\def\mbs   {\ensuremath{\mbox{\,Mb/s}}\xspace}
\def\gbs   {\ensuremath{\mbox{\,Gb/s}}\xspace}
\def\tbs   {\ensuremath{\mbox{\,Tb/s}}\xspace}

%% Time 
\def\sec  {\ensuremath{\rm {\,s}}\xspace}
\def\ms   {\ensuremath{{\rm \,ms}}\xspace}
\def\mus  {\rm \,\textmu s\xspace}
\def\ns   {\ensuremath{{\rm \,ns}}\xspace}
\def\ps   {\ensuremath{{\rm \,ps}}\xspace}
\def\fs   {\ensuremath{\rm \,fs}\xspace}

\def\mhz  {\ensuremath{{\rm \,MHz}}\xspace}
\def\khz  {\ensuremath{{\rm \,kHz}}\xspace}
\def\hz   {\ensuremath{{\rm \,Hz}}\xspace}

\def\invps{\ensuremath{{\rm \,ps^{-1}}}\xspace}

\def\yr   {\ensuremath{\rm \,yr}\xspace}
\def\hr   {\ensuremath{\rm \,hr}\xspace}
%[{"type":"track","uniqueId":"B07FMHZNT4","libraryId":"","context":"prime","asin":"B07FMHZNT4"}]
%% Temperature
\def\degc {\ensuremath{^\circ}{C}\xspace}
\def\degk {\ensuremath {\rm K}\xspace}

%% Material lengths, radiation
\def\Xrad {\ensuremath{X_0}\xspace}
\def\NIL{\ensuremath{\lambda_{int}}\xspace}
\def\mip {MIP\xspace}
\def\neutroneq {\ensuremath{\rm \,n_{eq}}\xspace}
\def\neqcmcm {\ensuremath{\rm \,n_{eq} / cm^2}\xspace}
\def\kRad {\ensuremath{\rm \,kRad}\xspace}
\def\MRad {\ensuremath{\rm \,MRad}\xspace}
\def\ci {\ensuremath{\rm \,Ci}\xspace}
\def\mci {\ensuremath{\rm \,mCi}\xspace}

%% Uncertainties
\def\sx    {\ensuremath{\sigma_x}\xspace}    
\def\sy    {\ensuremath{\sigma_y}\xspace}   
\def\sz    {\ensuremath{\sigma_z}\xspace}    

\newcommand{\stat}{\ensuremath{\mathrm{(stat)}}\xspace}
\newcommand{\syst}{\ensuremath{\mathrm{(syst)}}\xspace}
\newcommand{\model}{\ensuremath{\mathrm{(model)}}\xspace}

%% Maths
\newcommand{\ten}[1]{\ensuremath{\times 10^{#1}}}
\def\order{{\ensuremath{\cal O}}\xspace}
\newcommand{\chisq}{\ensuremath{\chi^2}\xspace}
\newcommand{\erfc}[1]{\ensuremath{\rm{Erfc}(#1)}\xspace}

\def\deriv {\ensuremath{\mathrm{d}}}

\def\gsim{{~\raise.15em\hbox{$>$}\kern-.85em
          \lower.35em\hbox{$\sim$}~}\xspace}
\def\lsim{{~\raise.15em\hbox{$<$}\kern-.85em
          \lower.35em\hbox{$\sim$}~}\xspace}

\newcommand{\DR}{\ensuremath{\Delta \text{R}}}
\newcommand{\DPhi}{\ensuremath{\Delta\phi}}
\newcommand{\DEta}{\ensuremath{\Delta\eta}}

\articletype{Topical Review} %	 e.g. Paper, Letter, Topical Review...

\title{Heavy Flavors and Quarkonia at RHIC}

%\author{Cameron Dean$^1$\orcid{0000-0002-6002-5870}, Xin Dong$^2$\orcid{0000-0001-9083-5906}, Rongrong Ma$^{3,*}$\orcid{0000-0002-3834-5026}, Cesar Luiz da Silva$^{4}$\orcid{0000-0003-4106-8258}}
\author{Cameron Dean$^1$, Xin Dong$^2$, Rongrong Ma$^{3,*}$, Cesar Luiz da Silva$^{4}$}

\affil{$^1$Massachusetts Institute of Technology, Cambridge, MA, United States}

\affil{$^2$Lawrence Berkeley National Laboratory, Berkeley, CA, United States}

\affil{$^3$Brookhaven National Laboratory, Upton, NY, United States}

\affil{$^4$Los Alamos National Laboratory, Los Alamos, NM, United States}

\affil{$^*$Author to whom any correspondence should be addressed.}

\email{marr@bnl.gov}


%\tableofcontents

\justifying

\begin{abstract}
\justifying
After the discovery of the strongly coupled quark-gluon plasma (QGP), a nearly perfect fluid, in 200 GeV Au+Au collisions at RHIC, understanding its microscopic structure and transport properties has become a central goal of relativistic heavy-ion physics. Heavy-flavor particles, containing charm or bottom quarks, provide unique sensitivity to the QGP because they are produced predominantly in the initial hard scatterings and interact with the medium throughout its evolution. This review summarizes measurements of open heavy flavor and quarkonia by the PHENIX and STAR experiments, focusing primarily on 200 GeV collisions recorded during the first two decades of RHIC operations. We discuss key results on charm, bottom, and quarkonium production cross sections; cold nuclear matter effects in small collision systems; nuclear modification factors and elliptic flow of open heavy-flavor hadrons; charm baryon-to-meson production ratios; and the suppression patterns of quarkonium states in Au+Au collisions. We further highlight the resulting insights into the properties of the QGP, including heavy-quark transport, hadronization mechanisms, and quarkonium dissociation and regeneration in the medium. Finally, we discuss the future prospects of the RHIC heavy-flavor program, enabled by the large data sets collected by the STAR and \sphenix experiments, and its strong synergy with the future Electron-Ion Collider, where precision measurements of heavy-flavor production in electron--proton and electron--nucleus collisions will provide complementary constraints on the structure of QCD matter.

\end{abstract}

\section{Introduction}
One of the central goals of modern nuclear physics is to understand the properties of strongly interacting matter under extreme temperatures and energy densities, and to elucidate its role in the evolution of the early universe~\cite{Yagi:2005yb}. Such conditions are recreated in heavy-ion collisions at the Relativistic Heavy Ion Collider (RHIC) and the Large Hadron Collider (LHC), where quarks and gluons are no long confined inside hadrons. The resulting state of deconfined matter, known as the quark--gluon plasma (QGP), has been established~\cite{STAR:2005gfr,PHENIX:2004vcz,PHOBOS:2004zne,BRAHMS:2004adc,ALICE:2022wpn,CMS:2024krd}. 

The strong interaction is described by Quantum Chromodynamics (QCD), whose non-perturbative many-body dynamics make quantitative calculations of the QGP particularly challenging. Experimental measurements using light- and strange-flavor hadrons have revealed strong collective flow and substantial parton energy loss, demonstrating that the QGP produced at RHIC and the LHC behaves as a nearly perfect liquid with exceptionally low viscosity~\cite{STAR:2005gfr,PHENIX:2004vcz,PHOBOS:2004zne,BRAHMS:2004adc,ALICE:2022wpn,CMS:2024krd}. Nevertheless, the microscopic properties of this strongly coupled QCD medium remain active areas of investigation.

Heavy flavor particles, which contain at least a bottom or charm quark, provide unique probes of the microscopic nature of this plasma. As their masses are above the QCD scale and the critical temperature for the QGP formation~\cite{ParticleDataGroup:2024cfk, STAR:2024bpc}, they are produced predominantly in the initial hard partonic scatterings and subsequently experience the entire evolution of the QGP. Two types of heavy flavor particles are widely used, quarkonia which are composed of a charm--anticharm or bottom--antibottom pair, and open heavy-flavor hadrons which contain a charm or bottom quark in addition to a quark of a different flavor~\footnote{The $B_c^+$ meson contains a charm and antibottom quark but is also classified as open bottom.}.

Because the masses of charm and bottom quarks are much larger than the characteristic temperature of the QGP and the masses of its constituents, their propagation through the medium can be approximated as Brownian motion before they hadronize into open heavy-flavor hadrons. Consequently, their interactions with the medium can be described within kinetic transport frameworks over a broad momentum range, providing valuable insight into the microscopic properties of the QGP at different scales. Measurements of the nuclear modification of open heavy-flavor hadrons, particularly in comparison with light-flavor hadrons, probe the mass dependence of parton energy loss. When combined with measurements of elliptic flow ($v_2$), they provide stringent constraints on the heavy-quark diffusion coefficient and transport properties of the medium~\cite{Moore:2004tg,Dong:2019byy,He:2022ywp}. Unlike light quarks, heavy quarks largely retain their flavor identity after being produced in the initial hard scatterings. Consequently, the relative production rates of different species provide unique sensitivity to heavy-quark hadronization mechanisms in both proton-proton (\pp) and heavy-ion (\AA) collisions. Examples include color reconnection, in which quarks originating from different parton showers reconnect before hadronization~\cite{Christiansen:2015yqa}, and quark coalescence in the hot QCD medium, where competing scenarios such as sequential and simultaneous coalescence predict distinct heavy-flavor hadrochemistry~\cite{Zhao:2018jlw}.

Quarkonia, color-neutral bound states, have long been regarded as unique probes of both cold and hot QCD matter. It was originally proposed that color screening in the QGP would weaken the binding between the heavy quark and antiquark, leading to quarkonium dissociation and consequently a suppression of quarkonium yields in heavy-ion collisions relative to those in \pp\ collisions~\cite{Matsui:1986dk,Satz:2005hx,Andronic:2025jbp}. Recent advances in lattice QCD have refined this picture~\cite{Bazavov:2023dci,Ding:2025fvo}. Rather than being governed solely by Debye screening of the binding potential, quarkonium suppression is now understood to arise primarily from in-medium dissociation induced by interactions with deconfined quarks and gluons. Since the dissociation rate increases with medium temperature and decreases with quarkonium binding energy, more weakly bound states are expected to dissociate at lower temperatures. This leads to the characteristic pattern of sequential suppression, in which excited quarkonium states exhibit progressively stronger suppression than their more tightly bound counterparts, providing a sensitive probe of the thermodynamic properties of the QGP.

In heavy-ion collisions, two of the most widely used observables are the nuclear modification factor (\RAA) and $v_2$. The nuclear modification factor is defined as the ratio of the particle yield measured in heavy-ion collisions ($dN_{\rm AA}/dp_{\rm T}$) to that in \pp\ collisions ($dN_{pp}/dp_{\rm T}$), scaled by the number of binary nucleon--nucleon collisions, \ncoll, as given in Eq.~\ref{eq:RAA}:
\begin{equation}
R_{\rm AA}(p_{\rm T}) \equiv \frac{dN_{\rm AA}/dp_{\rm T}}{N_{\rm coll}\times dN_{pp}/dp_{\rm T}}.
\label{eq:RAA}
\end{equation}
It quantifies the modification of particle production due to the nuclear environment. An \RAA\ value of unity indicates that particle production in heavy-ion collisions is consistent with an incoherent superposition of nucleon-nucleon collisions.

In non-central heavy-ion collisions, the initial spatial anisotropy of the overlap region is converted into momentum-space anisotropy through the pressure gradients developed during the hydrodynamic expansion of the QGP. The elliptic flow coefficient, $v_2$, is defined as the second Fourier coefficient of the azimuthal distribution of particle yields with respect to the event-plane angle, as given in Eq.~\ref{eq:v2}:
\begin{equation}
v_2 \equiv \left\langle \cos\left[2(\phi-\Psi)\right]\right\rangle,
\label{eq:v2}
\end{equation}
where $\phi$ is the azimuthal angle of the particle and $\Psi$ is the event-plane angle. The $v_2$ observable is sensitive to the early-time dynamics of the collision and the degree of collectivity developed in the QGP. It is therefore widely used to characterize the transport properties of the medium.

\section{Open Heavy Flavor}

\subsection{Signal reconstruction}
Open heavy-flavor hadrons have proper decay lengths ranging from a few tens to several hundreds of micrometers. Their precise reconstruction therefore requires high-resolution vertex detectors capable of resolving displaced decay vertices from the primary collision vertex. To meet this challenge, RHIC experiments developed and deployed state-of-the-art silicon vertex detectors. The STAR experiment pioneered the first large-scale application of a Monolithic Active Pixel Sensor (MAPS)-based silicon pixel detector, the Heavy Flavor Tracker (HFT)~\cite{Contin:2017mck}, while a second-generation MAPS-based vertex detector was subsequently implemented in the \sphenix experiment. The PHENIX experiment commissioned the Vertex Detector (VTX)~\cite{Ichimiya:2008am} and the Forward Vertex Detector (FVTX)~\cite{Aidala:2013vna}, which significantly enhanced its capabilities in measuring heavy-flavor hadrons at midrapidity and forward rapidity, respectively.

\subsection{\pp\ collisions}
\label{sec:charm_bottom_Xsec}

\begin{figure}
\centering
\includegraphics[width=0.48\textwidth]{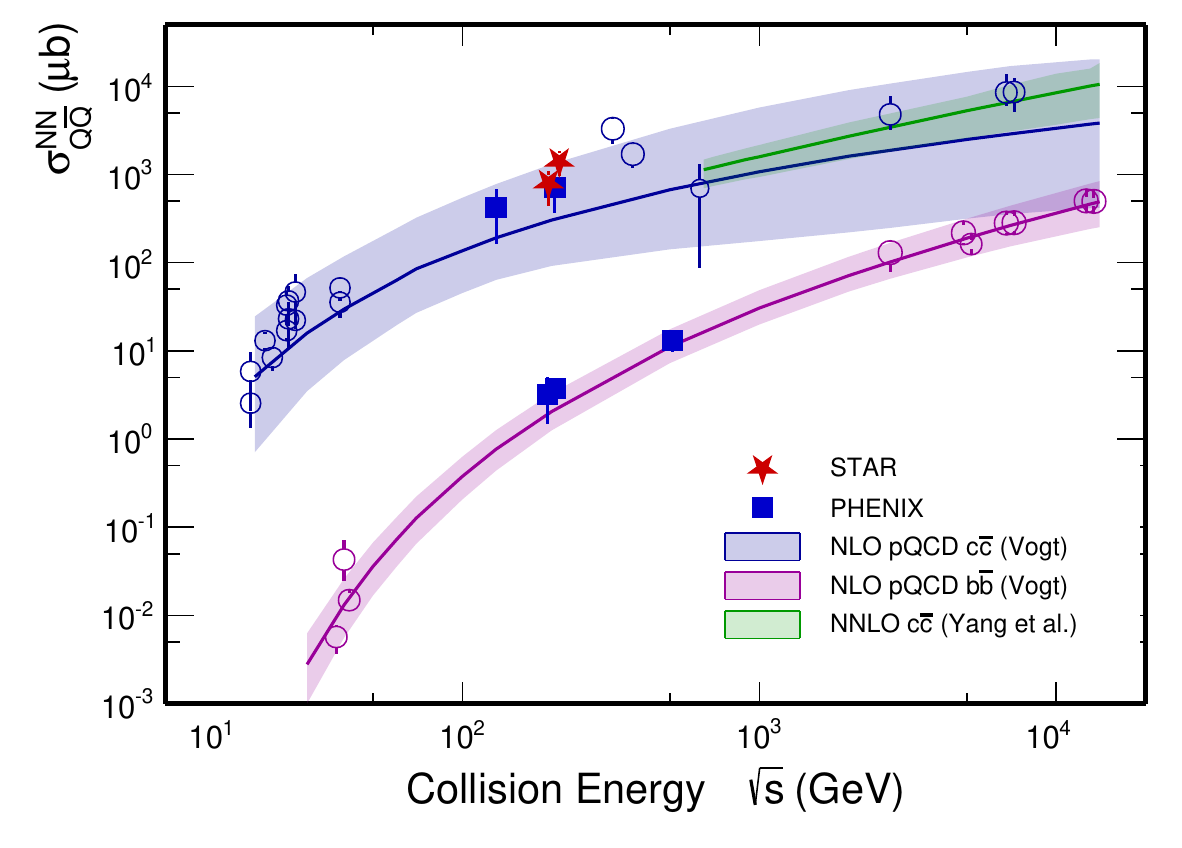}
\includegraphics[width=0.48\linewidth]{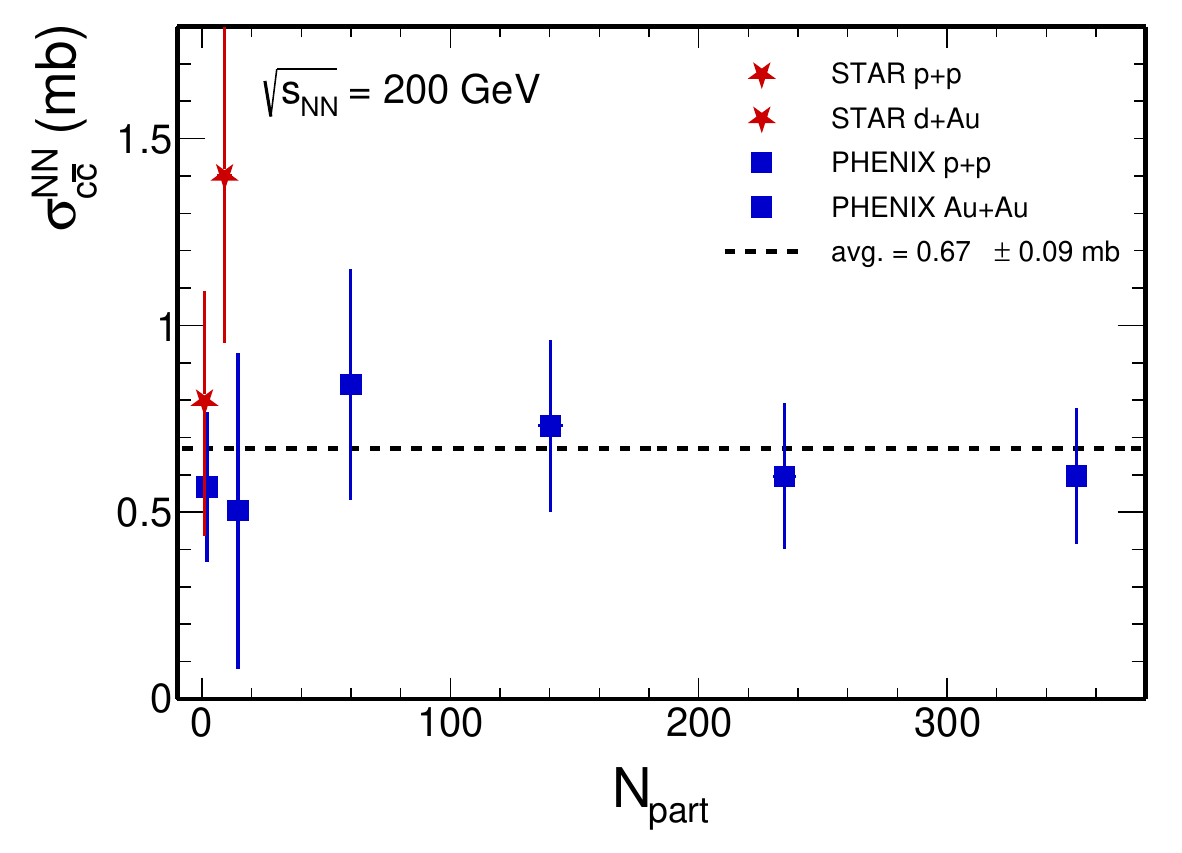}
\caption{Left: energy dependence of $c\bar{c}$ and $b\bar{b}$ production cross sections per nucleon-nucleon collisions measured in \pp, $p$+A and $d$+A collisions. Solid markers are from PHENIX and STAR measurements~\cite{PHENIX:2002ecc,PHENIX:2006tli,PHENIX:2013klq,PHENIX:2004ggw,PHENIX:2018dwt,PHENIX:2017exy,STAR:2004ocv,STAR:2012nbd}, while open circles are from other experiments~\cite{Bierlich:2023ewv,Jansen:1994bz,E771:1997rbx,HERA-B:2005tnp,ALICE:2014aev,ALICE:2012acz,ALICE:2021edd,ALICE:2021lmn,LHCb:2010wqx,LHCb:2015foc}. Color bands are NLO and NNLO pQCD calculations from Vogt~\cite{Vogt:2007aw} and Yang~\cite{Yang:2025pmq}. Right: PHENIX and STAR measurements of total $c\bar{c}$ production cross section per nucleon-nucleon collision in $p$+$p$, $d$+Au, and Au+Au collisions at \sqrtsnn = 200\,GeV~\cite{PHENIX:2006tli,PHENIX:2004ggw,STAR:2004ocv,STAR:2012nbd}.}
\label{fig:heavy_flavor_energy}
\end{figure}

Charm and bottom quark production cross sections in \pp\ collisions provide essential benchmarks for understanding heavy-quark production mechanisms and serve as baselines for interpreting their measurements in heavy-ion collisions. RHIC experiments employed a variety of complementary techniques to measure charm and bottom production over broad kinematic ranges~\cite{PHENIX:2002ecc,PHENIX:2006tli,PHENIX:2013klq,PHENIX:2004ggw,PHENIX:2018dwt,PHENIX:2017exy,STAR:2004ocv,STAR:2012nbd}. Total production cross sections are obtained by extrapolating the measured differential cross sections to zero \pt\ and full rapidity, typically using different PYTHIA tunings to estimate the unmeasured phase space. As shown in the left panel of Fig.~\ref{fig:heavy_flavor_energy}, the resulting charm and bottom cross sections measured in \pp\ and $d$+Au collisions are consistent with world data~\cite{Bierlich:2023ewv,Jansen:1994bz,E771:1997rbx,HERA-B:2005tnp,ALICE:2014aev,ALICE:2012acz,ALICE:2021edd,ALICE:2021lmn,LHCb:2010wqx,LHCb:2015foc} and with perturbative QCD calculations at next-to-leading order (NLO) and next-to-next-to-leading order (NNLO)~\cite{Vogt:2007aw,Yang:2025pmq}. The dominant uncertainties arise from the model-dependent extrapolations to the unmeasured kinematic regions.

The right panel of Fig.~\ref{fig:heavy_flavor_energy} shows the total charm production cross section per \ncoll\ as a function of collision centrality, from \pp\ to central \AuAu\ collisions at the center-of-mass energy per nucleon-nucleon pair (\sqrtsnn) of 200 GeV~\cite{PHENIX:2006tli,PHENIX:2004ggw,STAR:2004ocv,STAR:2012nbd}. Within uncertainties, the charm production cross section per binary collision is independent of collision centrality, demonstrating that charm quarks are produced predominantly in the initial hard partonic scatterings and that thermal charm production in the QGP is negligible at RHIC energies. For the 0--10\% most central \AuAu\ collisions, this corresponds to approximately 15 $c\bar{c}$ pairs produced per event, using $\langle N_{\mathrm{coll}}\rangle \approx 960$ from a Glauber-model calculation~\cite{Miller:2007ri}. \ncoll\ scaling of bottom production has also been observed by the PHENIX experiment through measurements of $B\rightarrow J/\psi$ decays in Cu+Au collisions~\cite{PHENIX:2017caf}, indicating that bottom quarks are likewise produced predominantly in the initial hard scatterings.

\begin{figure}
    \begin{minipage}{0.48\linewidth}
        \includegraphics[width=1.0\linewidth]{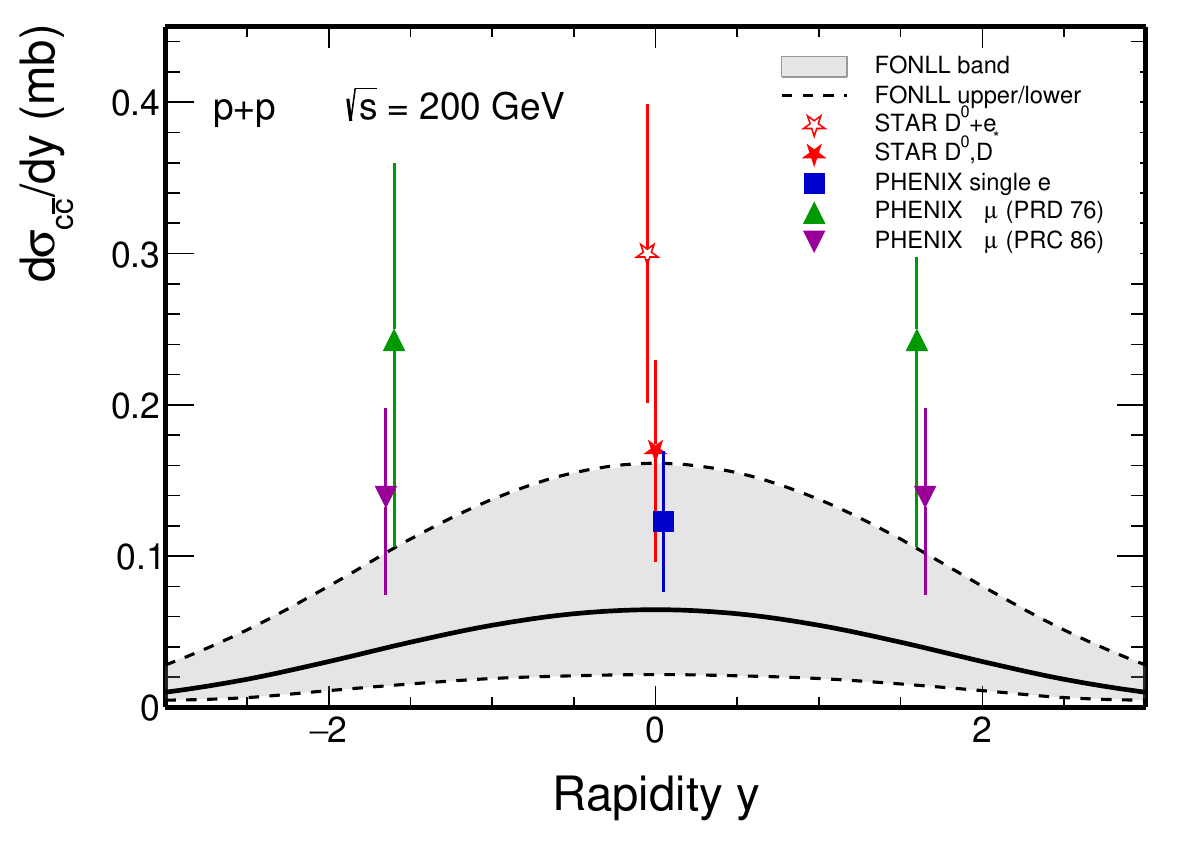}
    \end{minipage}
    \begin{minipage}{0.48\linewidth}
        \includegraphics[width=1.0\linewidth]{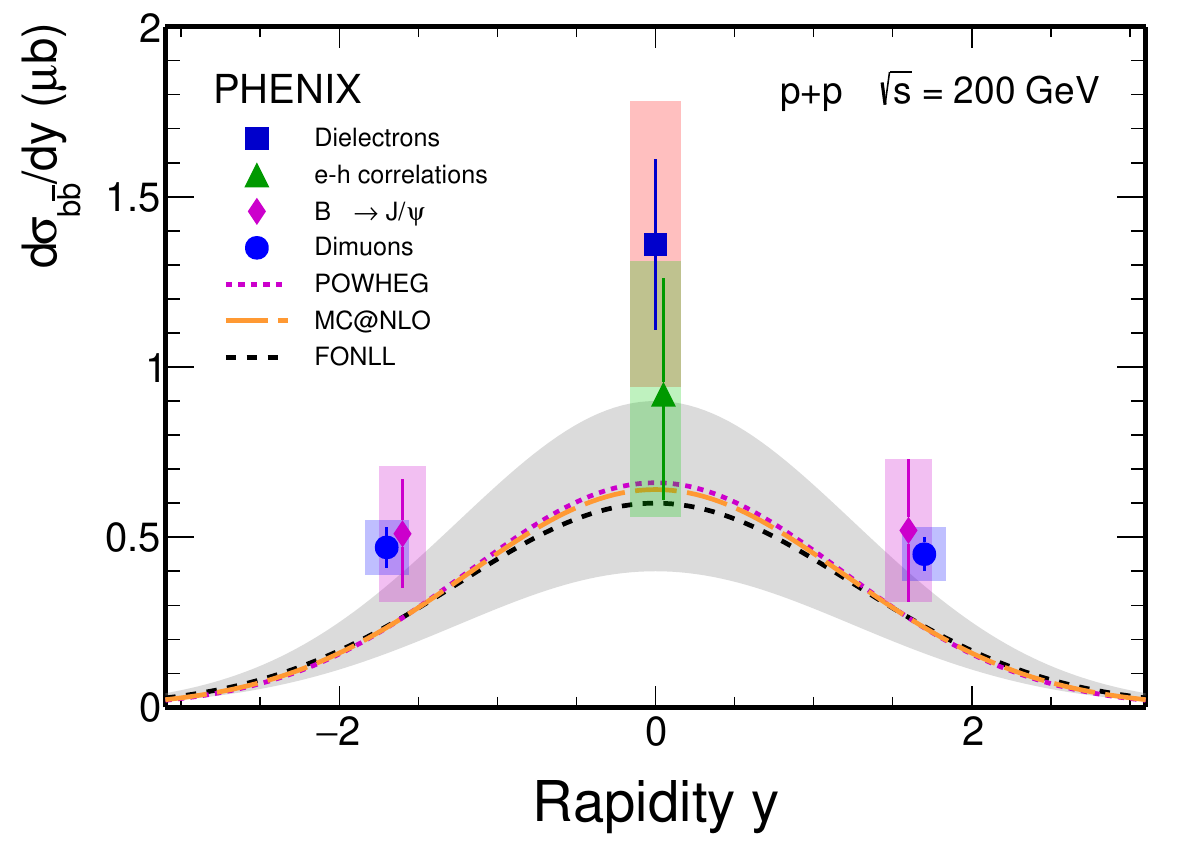}
    \end{minipage}
    \caption{Rapidity dependent charm and bottom differential cross sections obtained from different measurements~\cite{PHENIX:2004ggw,PHENIX:2018dwt,STAR:2004ocv,PHENIX:2008qav,PHENIX:2006fxr,Read:2012uz} and comparison to FONLL calculations~\cite{Cacciari:1998it,Cacciari:2001td,Cacciari:2012ny} and event generator predictions~\cite{Oleari:2010nx,Frixione:2002bd}.}
    \label{fig:dsigmaHF_dy}
\end{figure}

Figure~\ref{fig:dsigmaHF_dy} summarizes the rapidity dependence of the \pt-integrated charm and bottom production cross sections measured at RHIC~\cite{PHENIX:2004ggw,PHENIX:2018dwt,STAR:2004ocv,PHENIX:2008qav,PHENIX:2006fxr,Read:2012uz}, together with comparisons to  calculations at fixed-order plus next-to-leading logs (FONLL)~\cite{Cacciari:1998it,Cacciari:2001td,Cacciari:2012ny} and event generators~\cite{Oleari:2010nx,Frixione:2002bd}. Overall, the measured bottom production cross sections are approximately a factor of two larger than the central values of the theoretical predictions. However, they remain consistent within the sizable experimental and theoretical uncertainties.

Studies of angular correlations between dimuons from correlated $D\bar{D}$ and $B\bar{B}$ decays provide further insight into the underlying heavy-quark production mechanisms~\cite{PHENIX:2018dwt}. At RHIC energies, charm production receives contributions, in descending order, from flavor excitation, pair creation, and gluon splitting. In contrast, bottom production is dominated by pair creation and flavor excitation, while the contribution from gluon splitting is negligible. This production pattern differs from that at lower-energy fixed-target experiments, where pair creation dominates both charm and bottom production \cite{Lourenco:2006vw}, and from the Tevatron and LHC, where gluon splitting becomes increasingly important at higher collision energies~\cite{CDF:2004mmv,CMS:2011yuk,LHCb:2012aiv}. The dominance of pair creation and flavor excitation for bottom production at RHIC results in strong initial $b\bar{b}$ correlations, making RHIC uniquely sensitive to medium-induced modifications of correlated bottom-quark pairs in heavy-ion collisions. It also provides a cleaner experimental environment for reconstructing and interpreting bottom-hadron observables than at higher collision energies.

\subsection{Small system collisions}
\label{sec:CNM}

The presence of a heavy nucleus in a collision can modify heavy-flavor production even in the absence of a QGP. These modifications are collectively referred to as cold nuclear matter (CNM) effects. Experimentally, CNM effects are commonly quantified through measurements of nuclear modification factors in \pau\ and \dAu\ collisions as no QGP with extended volume is expected to form in such small-system collisions, allowing CNM effects to be isolated. By convention, the $p$ and $d$ beams at RHIC travel in the forward-rapidity direction, while the ion beam travels in the backward-rapidity direction.

\begin{figure}[htb]
    \centering
    \begin{minipage}{0.49\linewidth}
    \includegraphics[width=1\linewidth]{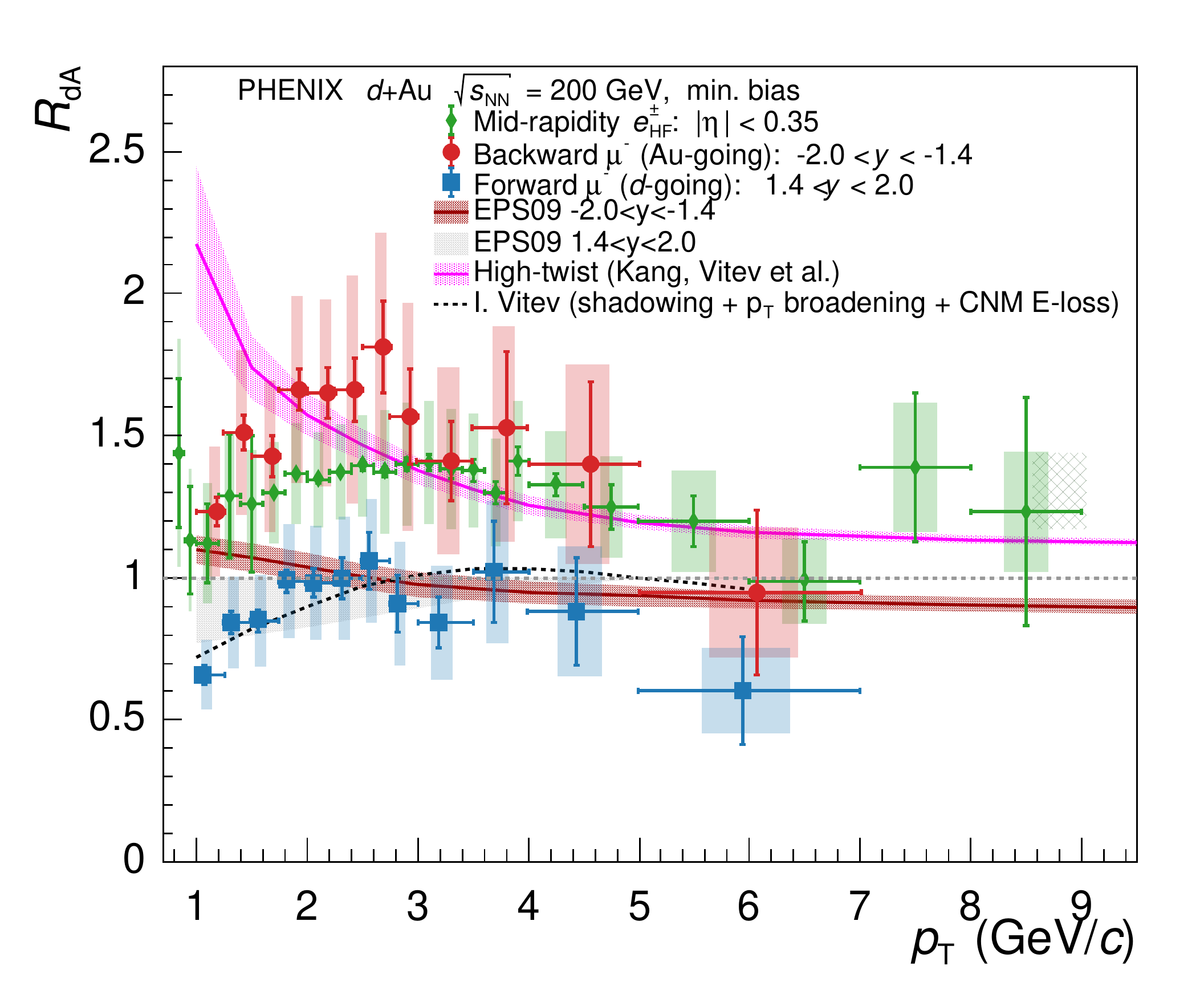}
    \end{minipage}
    \begin{minipage}{0.49\linewidth}
    \includegraphics[width=1.0\linewidth]{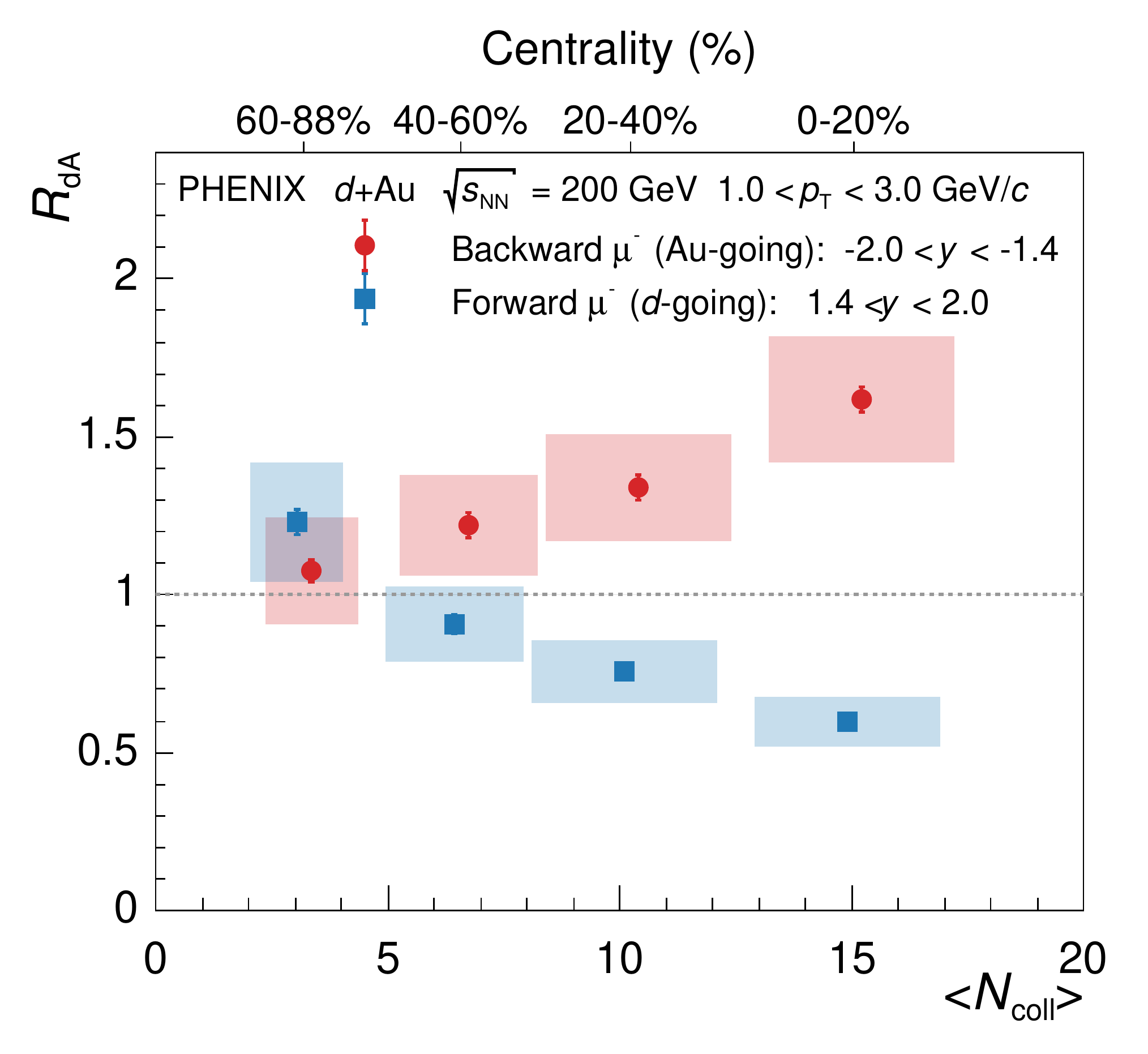}
    \end{minipage}
    \caption{\rda\ of leptons from heavy-flavor hadron decays in \dAu\ collisions at \sqrtsnn\ = 200 GeV as functions of \pt\ (left) and the average \ncoll\ (right), in three rapidity regions. Electrons are measured at midrapidity~\cite{PHENIX:2012hww}, while muons are measured at forward and backward rapidities~\cite{PHENIX:2013txu}. Data are compared to calculations based on the EPS09 nPDFs~\cite{Eskola:2009uj}, a pQCD high-twist calculation for backward rapidity~\cite{Kang:2014hha}, and calculations incorporating initial-state effects for forward rapidity~\cite{Vitev:2007ve}.}

    \label{fig:HF_RdA}
\end{figure}

At RHIC, CNM effects on heavy-flavor production have been studied using leptons from charm- and bottom-hadron decays measured in \pp\ and \dAu\ collisions~\cite{PHENIX:2012hww,PHENIX:2013txu}. The resulting nuclear modification factor, \rda, is shown in Fig.~\ref{fig:HF_RdA} for three rapidity intervals. At forward rapidity, heavy-flavor production predominantly probes gluons in the Au nucleus with Bjorken-$x$ values of \mbox{$3\times10^{-3}<x_{\rm Au}<10^{-2}$}, where a suppression is observed. This behavior is broadly consistent with expectations from modifications to nuclear parton distribution functions (nPDFs) compared to those in free nucleons~\cite{Eskola:2009uj} or initial-state effects including shadowing, initial-state parton energy loss, isospin effects, and \pt\ broadening~\cite{Vitev:2007ve}. At mid- and backward rapidities, corresponding to \mbox{$10^{-2}<x_{\rm Au}<10^{-1}$} and \mbox{$0.1<x_{\rm Au}<0.3$}, respectively, heavy-flavor production probes the antishadowing region of nPDFs. In both regions, an enhancement of heavy-flavor yields is observed in \dAu\ collisions compared to that in \pp\ collisions. However, the magnitude of the enhancement exceeds that expected from nPDF modifications alone. Perturbative QCD calculations that additionally incorporate incoherent multiple scattering reproduce the observed trend qualitatively~\cite{Kang:2014hha}, indicating that multiple CNM mechanisms contribute to the measured nuclear modification.

The right panel of Fig.~\ref{fig:HF_RdA} further demonstrates a pronounced dependence of \rda\ on average \ncoll\ or the collision centrality. In peripheral collisions, \rda\ is consistent with unity, whereas significant suppression (enhancement) is observed in central collisions at forward (backward) rapidity. These measurements indicate that CNM effects could be dependent on the collision geometry, or equivalently the impact parameter. It should be noted that a 10--20\% bias in the determination of $\langle N_{\rm coll}\rangle$ for events containing a hard scattering has been identified for $\langle N_{\rm coll}\rangle<5$ through comparisons of high-\pt\ $\pi^0$ and direct-photon production~\cite{PHENIX:2023dxl}. This bias is smaller than the systematic uncertainties of the measurements shown in the right panel of Fig.~\ref{fig:HF_RdA}.

\subsection{Heavy-ion collisions}
\subsubsection{Heavy quark energy loss}

During the first decade of RHIC operations, the PHENIX and STAR experiments investigated heavy-flavor production in heavy-ion collisions primarily through semileptonic decays to electrons. Early theoretical expectations predicted substantially weaker suppression of heavy quarks than that of light-flavor partons due to the ``dead-cone'' effect, which suppresses small-angle gluon radiations by massive quarks~\cite{Dokshitzer:2001zm,Zhang:2003wk}. Surprisingly, measurements in central \AuAu\ collisions showed that \raa\ of heavy-flavor decayed electrons is suppressed to a level comparable to that of light hadrons~\cite{PHENIX:2004ggw,STAR:2006btx}, in striking contrast to the enhancement observed in $d$+Au collisions at the same rapidity. These observations demonstrated that radiative energy loss alone is insufficient to describe heavy-quark interactions with the QGP and established collisional energy loss as an essential component of heavy-quark transport in the medium~\cite{Moore:2004tg}.

\begin{figure}[htb]
    \centering
    \includegraphics[width=0.48\linewidth]{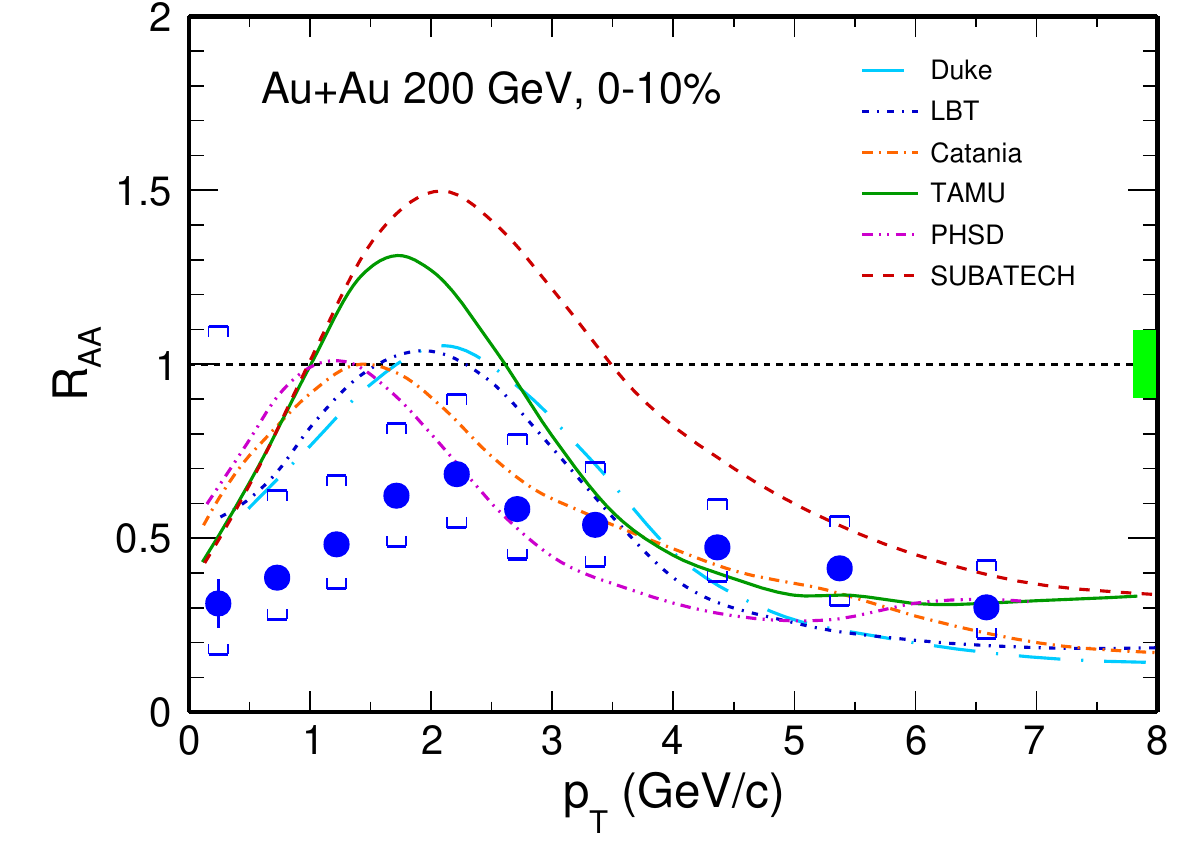}    \includegraphics[width=0.48\linewidth]{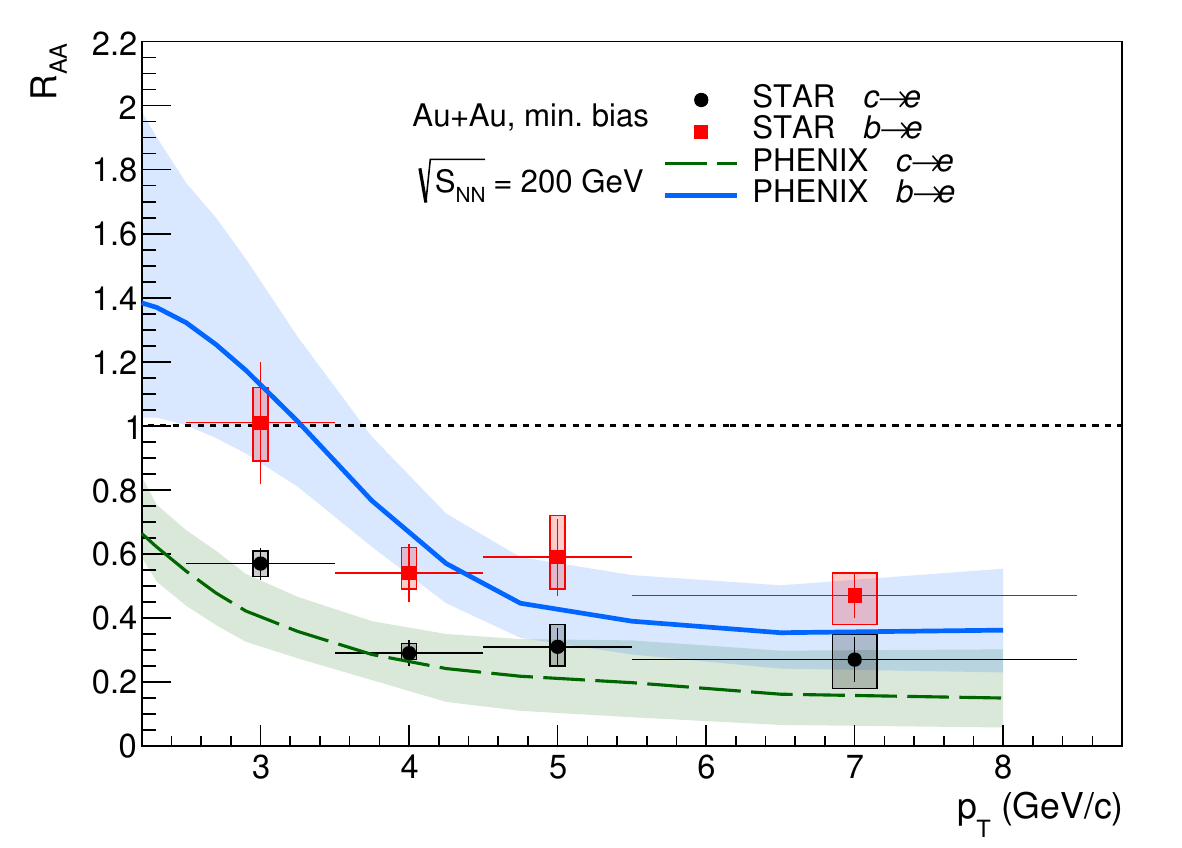}
    \caption{\raa\ of inclusive $D^0$ mesons in 0--10\% central \AuAu\ collisions (left)\cite{Adam:2018inb} and electrons from charm- and bottom-hadron decays in minimum-bias \AuAu\ collisions (right)~\cite{PHENIX:2015ynp,PHENIX:2022wim,STAR:2021uzu,STAR:2023qfk} at \sqrtsnn\ = 200 GeV. $D^0$-meson \raa\ is compared to various model calculations: Duke~\cite{Cao:2015hia}, LBT (Linearized Boltzmann Transport)~\cite{Cao:2016gvr}, Catania~\cite{Plumari:2019hzp}, TAMU~\cite{He:2019vgs}, PHSD (Parton-Hadron-String Dynamics)~\cite{Song:2015sfa}, and SUBATECH (MC@sHQ)~\cite{Nahrgang:2013xaa}.}
    \label{fig:rhic_raa}
\end{figure}

Because electrons from heavy-flavor decays receive contributions from both charm- and bottom-hadron decays, their interpretation is inherently model dependent. During the second decade of RHIC operations, both PHENIX and STAR installed precision silicon vertex detectors that enabled the statistical separation of electrons from charm- and bottom-hadron decays and, more importantly, the topological reconstruction of open-charm hadrons through their hadronic decay channels. As shown in the left panel of Fig.~\ref{fig:rhic_raa}, the $D^0$ meson \raa ~\cite{Adam:2018inb} exhibits strong suppression for \mbox{$p_{\rm T}\gtrsim5$ \gev}, indicating substantial energy loss of charm quarks as they traverse the QGP. This observation confirms the conclusion drawn from the earlier measurements of heavy-flavor decayed leptons. At low \pt, the $D^0$-meson \raa\ exhibits a characteristic ``bump'' structure that is qualitatively reproduced by a wide range of theoretical models~\cite{Nahrgang:2013xaa,He:2019vgs,Cao:2015hia,Cao:2016gvr,Song:2015sfa,Ke:2018jem,Li:2019lex,Plumari:2019hzp,Xu:2017obm}. This enhancement is generally attributed to the collective radial flow acquired by charm quarks through their interactions with the expanding QGP, together with hadronization via coalescence with flowing light quarks~\cite{Liu:2018syc}.

To investigate the mass dependence of parton energy loss, an indirect probe of bottom-quark energy loss is provided by electrons from bottom-hadron decays, identified through their displaced decay vertices. The PHENIX and STAR measurements of \raa\ for these electrons are shown in the right panel of Fig.~\ref{fig:rhic_raa}~\cite{PHENIX:2022wim,STAR:2021uzu}. Within the current experimental uncertainties, electrons from bottom-hadron decays exhibit systematically weaker suppression than those from charm-hadron decays, consistent with the expectation that heavier quarks lose less energy while traversing the QGP. However, the present uncertainties remain too large to establish this hierarchy quantitatively, underscoring the need for high-precision measurements of open-bottom production with the complete RHIC data sets.

\subsubsection{Charm hadron elliptic flow}

RHIC experiments investigated the collective behavior of heavy quarks through electrons from heavy-flavor hadron decays during the first decade of RHIC operations as well~\cite{PHENIX:2006iih,PHENIX:2010xji,STAR:2014yia}. These measurements provided the first indications of a sizable heavy-flavor $v_2$ in non-central \AuAu\ collisions. However, the experimental uncertainties were too large to place stringent constraints on the heavy-quark diffusion coefficient or to determine the degree of charm-quark thermalization in the QGP. Achieving these goals requires precision measurements of identified charm hadrons, particularly at low \pt, where collective flow effects are most pronounced.

\begin{figure}[htbp]
    \centering
    \includegraphics[width=0.5\linewidth]{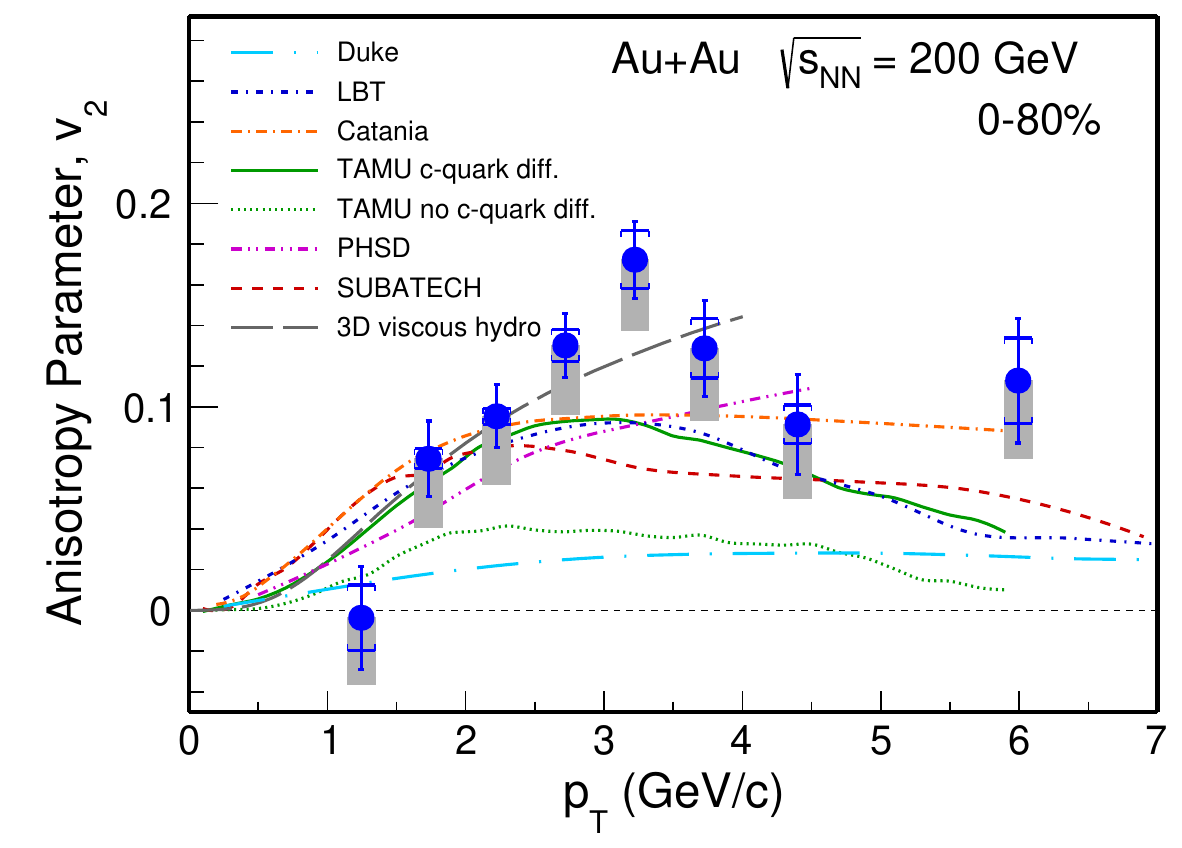}
    \caption{$D^0$ $v_2$ in 0-80\% minimum bias Au+Au collisions at $\sqrt{s_{\rm NN}}$ = 200 GeV~\cite{Adamczyk:2017xur} in comparison with various model calculations: Duke~\cite{Cao:2015hia}, LBT (Linearized Boltzmann Transport)~\cite{Cao:2016gvr}, Catania~\cite{Plumari:2019hzp}, TAMU ~\cite{He:2019vgs}, PHSD (Parton-Hadron-String Dynamics)~\cite{Song:2015sfa}, SUBATECH (MC@sHQ)~\cite{Nahrgang:2013xaa}, and 3D viscous hydrodynamics~\cite{Pang:2014ipa,Pang:2012he}.}
    \label{fig:d0_raa_v2}
\end{figure}

The installation of the HFT enabled STAR to perform precision measurements of the $D^0$-meson elliptic flow down to low \pt. Figure~\ref{fig:d0_raa_v2} presents the $D^0$-meson $v_2$ measured in 0--80\% minimum-bias \AuAu\ collisions~\cite{Adamczyk:2017xur}. The data suggest that charm quarks have accumulated sufficient scatterings and may have reached thermal equilibrium in the QGP medium. Furthermore, the measured $D^0$ $v_2$ approximately follows the empirical number-of-constituent-quark (NCQ) scaling~\cite{Molnar:2003ff,STAR:2003wqp} established for light- and strange-flavor hadrons, providing evidence that charm quarks develop substantial collectivity at the partonic stage. 

Also shown in Fig.~\ref{fig:d0_raa_v2} are comparisons with representative theoretical calculations~\cite{Nahrgang:2013xaa,He:2019vgs,Cao:2015hia,Cao:2016gvr,Song:2015sfa,Ke:2018jem,Li:2019lex,Plumari:2019hzp,Xu:2017obm}, most of which incorporate heavy-quark transport in the QGP. In particular, the TAMU calculation without charm-quark diffusion \cite{He:2019vgs} significantly underestimates the measured $D^0$ $v_2$, highlighting the importance of strong charm-medium interactions. In addition, calculations based on three-dimensional viscous hydrodynamics provide a good description of the data below $p_{\rm T}\approx4$~GeV/$c$ \cite{Pang:2012he,Pang:2014ipa}. With the implementation of the FVTX detector, PHENIX was able to measure heavy flavor in Au+Au collisions at forward rapidity ($1.2<|y|<2$), where the observed heavy-flavor $v_2$ is similar to mid-rapidity~\cite{PHENIX:2024bos}, indicating that the strong coupling of charm quarks to the medium extends over a broad rapidity range.

\subsubsection{Charm quark hadronization}

As heavy quarks are produced predominantly in the initial hard partonic scatterings and retain their flavor identity throughout the evolution of the QGP, they are unique probes of hadronization mechanisms. In the absence of nuclear effects, heavy-quark fragmentation is commonly described within the framework of collinear factorization, where fragmentation functions are primarily constrained by measurements in $e^+e^-$ and $ep$ collisions at LEP \cite{Cacciari:2005uk} and HERA \cite{ZEUS:2013xva,Bracinik:2006xx}.

Measurements of light- and strange-flavor hadrons have revealed a significant enhancement of baryon-to-meson production ratios in central heavy-ion collisions relative to \pp\ and peripheral heavy-ion collisions~\cite{STAR:2006uve,STAR:2011fbd}. Such enhancements are qualitatively described by hadronization models incorporating quark coalescence, in which nearby quarks in phase space recombine to form hadrons~\cite{Greco:2003xt}. Similar enhancements are predicted for charm hadrons, making charm baryon-to-meson ratios particularly sensitive to heavy-quark hadronization in the QGP~\cite{Oh:2009zj}. In addition, the enhanced abundance of strange quarks in the QGP is expected to increase the production of charm-strange mesons in heavy-ion collisions~\cite{He:2012df}.

\begin{figure}
    \centering
    \includegraphics[width=0.48\linewidth]{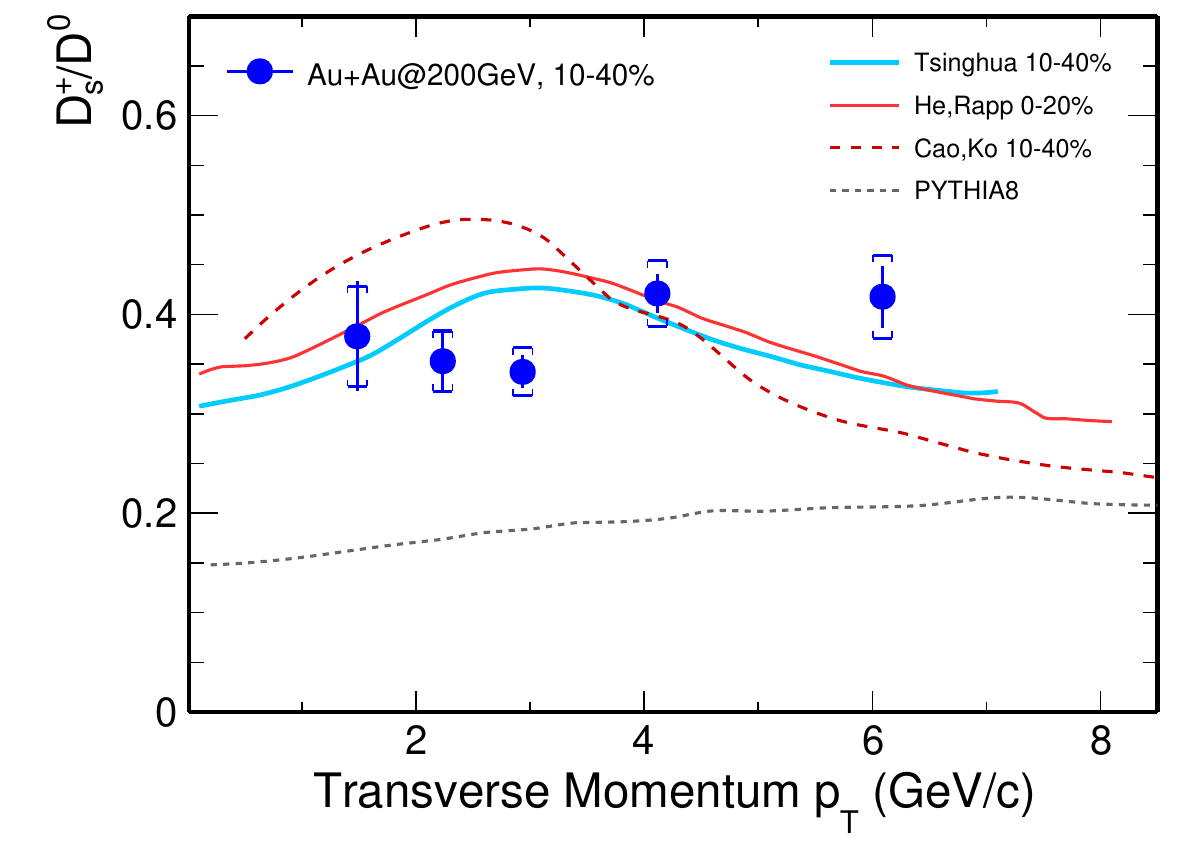}
    \includegraphics[width=0.48\linewidth]{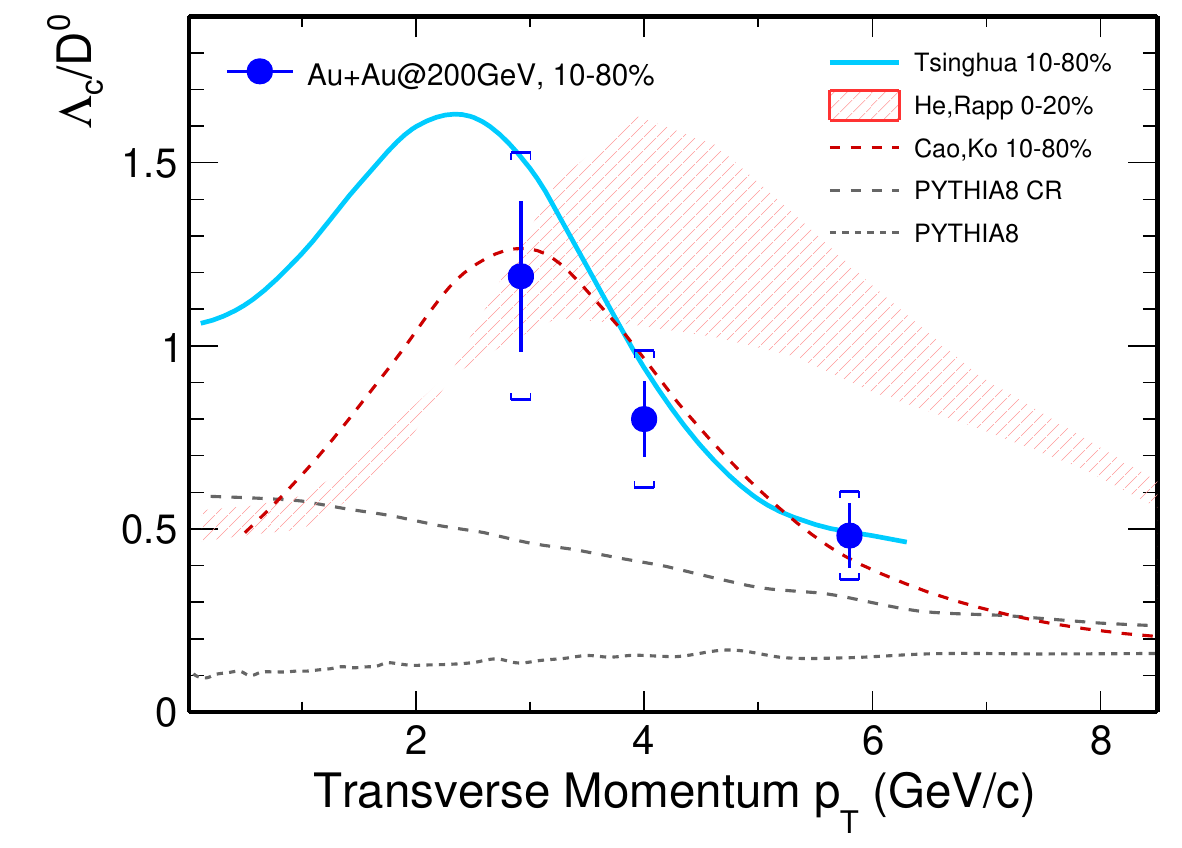}
    \caption{STAR measurements of $D_s^+ / D^0$ (left) and $\Lambda_c^+/D^0$ (right) ratios as a function of \pt in mid-central Au+Au collisions at \sqrtsnn\ = 200 GeV ~\cite{STAR:2021dsd0,STAR:2019ank} in comparisons with PYTHIA calculations \cite{Sjostrand:2007gs} and various model calculations incorporating coalescence hadronization for charm quarks~\cite{He:2019vgs,Plumari:2017ntm,Scardina:2017ipo}. }
    \label{fig:rhic_open_charm_ratios}
\end{figure}

STAR measurements of $D_s^+/D^0$~\cite{STAR:2021dsd0} and $\Lambda_c^+/D^0$~\cite{STAR:2019ank} production ratios in mid-central \AuAu\ collisions at \sqrtsnn\ = 200 GeV exhibit significant enhancements relative to PYTHIA predictions based on vacuum fragmentation, as shown in Fig.~\ref{fig:rhic_open_charm_ratios}. These observations are qualitatively consistent with the expectation of charm-quark hadronization through coalescence in the QGP, and are reproduced by several transport models incorporating coalescence hadronization~\cite{He:2019vgs,Plumari:2017ntm,Scardina:2017ipo}. For the $\Lambda_c^+/D^0$ ratio, early PYTHIA calculations based on fragmentation fractions extracted from $e^+e^-$ collisions substantially underestimate the measured enhancement \cite{Skands:2014pea}. More recent PYTHIA developments incorporating color reconnection and baryon junctions predict significantly larger $\Lambda_c^+/D^0$ ratios and successfully reproduce the \pp\ measurements at the LHC~\cite{Christiansen:2015yqa,ALICE:2020wfu}. However, precision measurements of $D_s^+/D^0$ and $\Lambda_c^+/D^0$ production ratios in \pp\ collisions at \sqrts\ = 200 GeV remain unavailable at RHIC. Such measurements are essential for establishing a data-driven baseline and disentangling vacuum hadronization effects from medium-induced modifications in heavy-ion collisions.

\subsubsection{Heavy quark diffusion coefficient}
Based on the Brownian-motion framework, heavy flavor interactions with the medium are commonly characterized by the heavy-quark spatial diffusion coefficient, ${\cal D}_s$\footnote{This notation should not be confused with the charm-strange meson $D_s$.}. A smaller value of ${\cal D}_s$ corresponds to stronger coupling between heavy quarks and the QGP, leading to more efficient momentum diffusion and energy loss. For comparisons among theoretical calculations, the dimensionless quantity $2\pi{\cal D}_sT$, where $T$ is the QGP temperature, is conventionally used.

\begin{figure}[htb]
    \centering
    \includegraphics[width=0.5\textwidth]{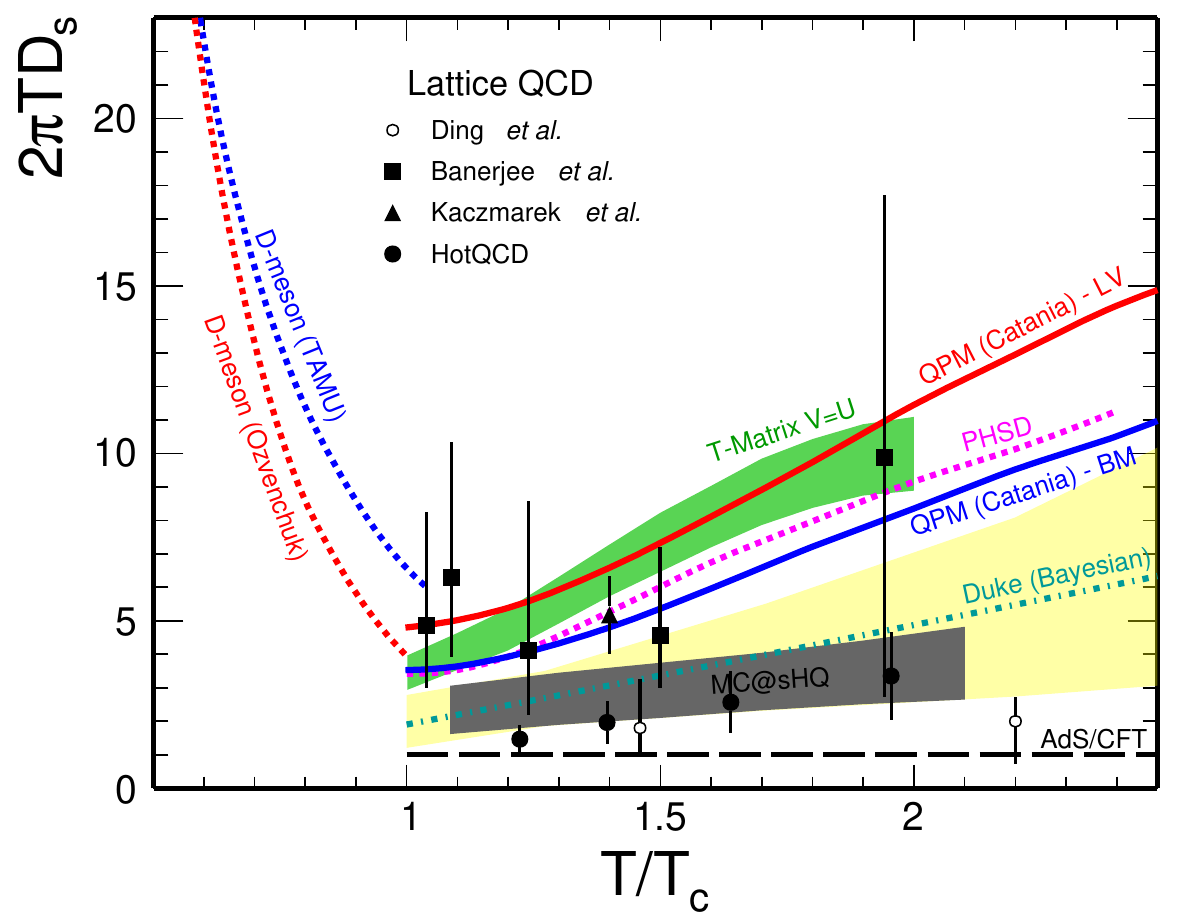}
    \caption{Dimensionless spatial diffusion coefficient $2\pi{\cal D}_sT$ obtained from lattice QCD~\cite{Altenkort:2023oms,Altenkort:2023eav,HotQCD:2025fbd}, AdS/CFT \cite{Casalderrey-Solana:2006fio}, and transport models, including MC@sHQ \cite{Nahrgang:2013xaa}, TAMU \cite{He:2019vgs},  Duke \cite{Cao:2015hia}, PHSD \cite{Song:2015sfa}, Catania \cite{Plumari:2019hzp}.
    }
    \label{fig:diffusion_coefficient}
\end{figure}

State-of-the-art lattice QCD calculations by the HotQCD Collaboration~\cite{Altenkort:2023oms,Altenkort:2023eav,HotQCD:2025fbd} predict values of $2\pi{\cal D}_sT$ approaching the quantum lower bound suggested by gauge/gravity duality (AdS/CFT)~\cite{Casalderrey-Solana:2006fio}, indicating strong interactions between charm quarks and the QGP. A variety of transport models, including the TAMU~\cite{He:2019vgs}, PHSD~\cite{Song:2015sfa}, Catania (QPM)~\cite{Plumari:2019hzp}, and MC@sHQ (SUBATECH)~\cite{Nahrgang:2013xaa} frameworks, incorporate different descriptions of heavy-quark interactions and hadronization in the medium. Despite these differences, they all provide a reasonable description of the measured $D^0$-meson \raa\ and $v_2$ at RHIC and the LHC. Figure~\ref{fig:diffusion_coefficient} summarizes the corresponding extractions of $2\pi{\cal D}_sT$, including Bayesian constraints from the Duke framework~\cite{Cao:2015hia} obtained through simultaneous comparisons to experimental \raa\ and $v_2$ measurements.

Remarkably, the diffusion coefficients extracted from these transport models are consistent with the lattice QCD calculations, all around 1--5 near the pseudo-critical temperature ($T_{\rm c}$). Their values are substantially smaller than those expected from perturbative QCD ~\cite{Moore:2004tg,Akiba:2015jwa}, providing compelling evidence that heavy-quark transport in the QGP is governed by strong, non-perturbative interactions. The corresponding scattering rates are estimated to be of the order of 0.5--1~GeV near $T_{\rm c}$~\cite{Dong:2019byy}, suggesting that well-defined quasiparticle excitations are strongly modified, or even dissolve, in this temperature regime. Together, these results establish heavy-quark diffusion as one of the most powerful quantitative probes of the microscopic transport properties of the QGP.

\subsubsection{Other developments}
While measurements of the $D^0$-meson \raa\ and $v_2$ have placed stringent constraints on the heavy-quark spatial diffusion coefficient near $T_{\rm c}$, its temperature dependence remains poorly constrained. The directed flow ($v_1$) of charm hadrons has been proposed as a complementary observable that is sensitive to the early-time dynamics of heavy quarks and the temperature dependence of the diffusion coefficient~\cite{Chatterjee:2017ahy}. The STAR measurement of the $D^0$-meson $v_1$ in \AuAu\ collisions revealed a sizable signal, significantly larger than those observed for light- and strange-flavor hadrons~\cite{STAR:2019clv}. This large directed flow is interpreted as evidence of the strong coupling between charm quarks and the initially tilted QGP medium. Existing transport model calculations \cite{Das:2016cwd,Chatterjee:2018lsx} generally underestimate the measured magnitude, indicating that the $D^0$ $v_1$ provides additional constraints on the temperature dependence of the heavy-quark diffusion coefficient beyond those obtained from \raa\ and $v_2$. The measured difference between the $D^0$ and $\overline{D}^0$ $v_1$ is consistent with zero within the current experimental uncertainties. Future high-statistics measurements of the $D$--$\overline{D}$ $v_1$ difference will provide a sensitive probe of electromagnetic fields in heavy-ion collisions~\cite{Das:2016cwd,Chatterjee:2018lsx}.

\subsection{Summary}

\begin{itemize}
\item The total charm and bottom production cross sections measured in \pp\ collisions at RHIC are generally consistent with perturbative QCD calculations within uncertainties. In \AuAu\ collisions, both charm and bottom production exhibit approximate binary-collision scaling, demonstrating that heavy quarks are produced predominantly in the initial hard scatterings with negligible thermal production in the QGP at RHIC energies. 

\item Cold nuclear matter (CNM) effects play a significant role in heavy-flavor production, leading to suppression at forward rapidity and enhancements at mid- and backward rapidities, which cannot be explained by nPDF effects alone, pointing to an interplay of multiple initial-state mechanisms. The observed dependence on collision centrality indicates a nontrivial impact-parameter dependence of CNM effects. 

\item Charm quarks undergo substantial energy loss while traversing the QGP, as demonstrated by the strong suppression of high-\pt\ $D^0$ mesons in central \AuAu\ collisions. Measurements of electrons from charm- and bottom-hadron decays further provide evidence for the expected mass hierarchy of parton energy loss, with bottom quarks exhibiting weaker suppression than charm quarks. 

\item The large heavy flavor $v_2$ demonstrates that charm quarks strongly couple to the QGP and participate in its collective expansion over a broad rapidity range. Transport models constrained by the measured $D^0$ \raa\ and $v_2$ favor a heavy-quark spatial diffusion coefficient of $2\pi{\cal D}_sT\approx1$--5 near the pseudo-critical temperature, consistent with state-of-the-art lattice QCD calculations. 

\item The enhanced $D_s^+/D^0$ and $\Lambda_c^+/D^0$ yield ratios observed in \AuAu\ collisions compared to PYTHIA predictions provide strong evidence that charm-quark hadronization is significantly modified by the QGP through quark coalescence. 
\end{itemize}

\section{Heavy Quarkonia}
Heavy quarkonia production involves both perturbative and non-perturbative processes. The initial creation of heavy quark–antiquark ($Q\bar{Q}$) pairs can be described with pQCD owing to the large heavy-quark mass. In contrast, the subsequent formation of a bound quarkonium state from the $Q\bar{Q}$ pair is a non-perturbative hadronization process, which requires experimental inputs to constrain theoretical models~\cite{Andronic:2015wma,Lansberg:2019adr}.

At RHIC, inclusive quarkonium production is measured predominantly as the experimental capability to separate prompt and non-prompt contributions is limited. The prompt component includes both directly produced quarkonia and those originating from feed-down decays of higher excited states. In the case of charmonium, the measured yield can also contain a non-prompt contribution from decays of bottom hadrons~\cite{Andronic:2015wma,Lansberg:2019adr}. These different sources complicate the interpretation of inclusive quarkonia measurements in heavy-ion collisions. Feed-down from excited states is particularly important because such states are generally more weakly bound and therefore more susceptible to in-medium dissociation, leading to an indirect suppression of the ground state. In addition, non-prompt quarkonia from bottom-hadron decays reflect the in-medium energy loss and transport of bottom quarks rather than the direct modification of quarkonium binding in the QGP. Consequently, a quantitative understanding of inclusive \RAA\ requires careful consideration of feed-down fractions and non-prompt contributions, as well as their respective medium modifications.

\subsection{Signal reconstruction}
Quarkonia decay promptly and are commonly reconstructed through their dilepton decay channels ($e^+e^-$ or $\mu^+\mu^-$), which can be efficiently triggered and identified using electromagnetic calorimeters or muon detectors. Precise and efficient identification of these decay leptons is essential for high-precision quarkonium measurements. The FVTX detector, installed in the PHENIX muon arms, played a crucial role in the observation of the $\psi(2S)$ state by significantly improving the dimuon opening-angle resolution, thereby enhancing the significance of the $\psi(2S)$ signal.

\subsection{\pp\ collisions}
\begin{figure}
    \centering
    \includegraphics[width=0.45\linewidth]{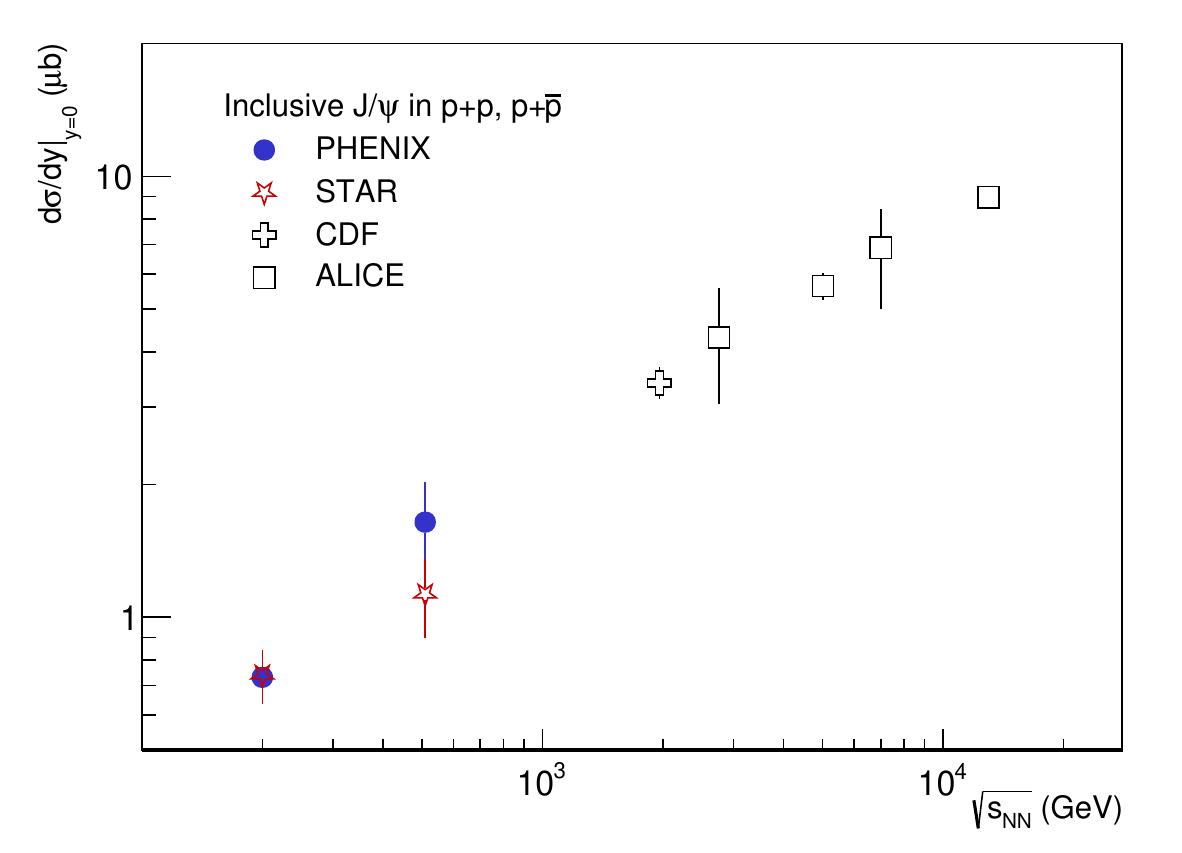}
    \includegraphics[width=0.45\linewidth]{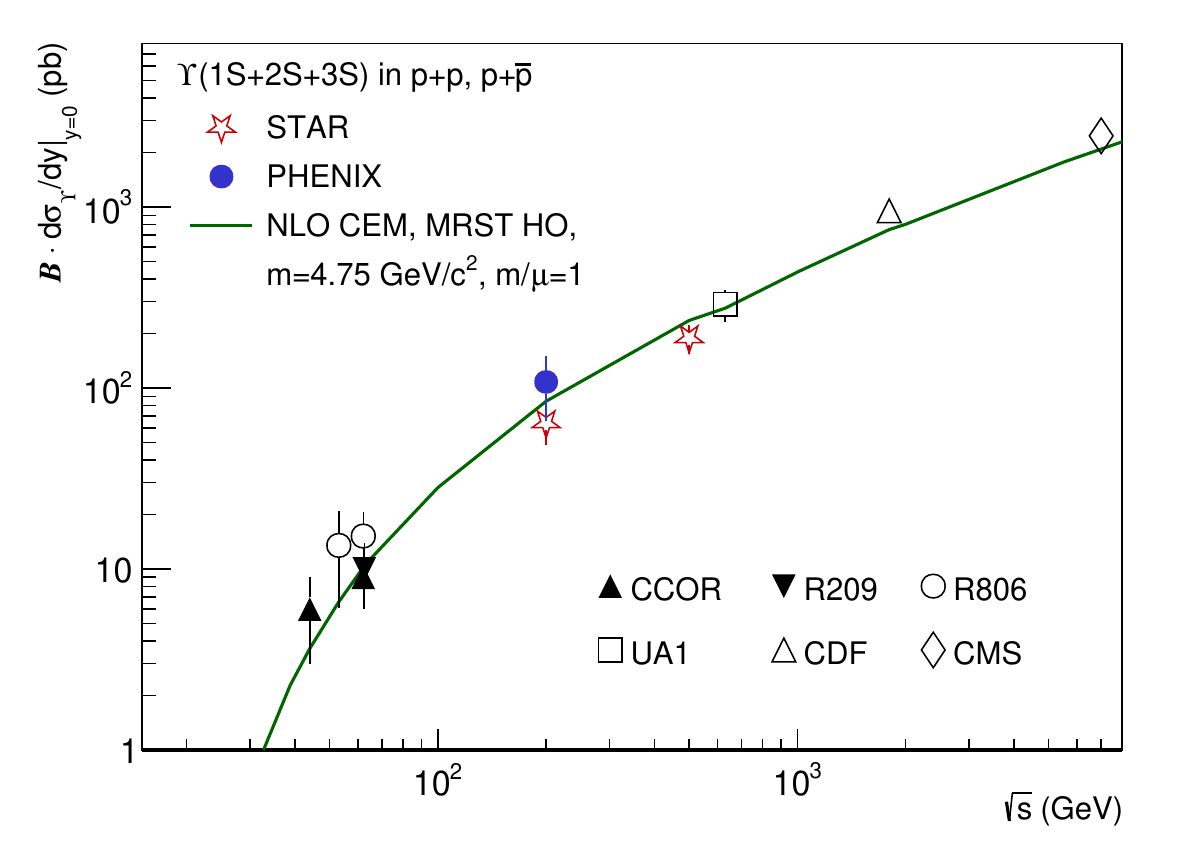}
    \caption{Inclusive \jpsi~\cite{PHENIX:2020dqu,PHENIX:2006aub,STAR:2019vkt,STAR:2021zvb,CDF:2004jtw,ALICE:2021dtt,ALICE:2019pid,ALICE:2012vup,ALICE:2011vrm} and \ups~\cite{STAR:2013kwk,STAR:2025ywy,PHENIX:2014tbe,CCOR:1979gxw,Kourkoumelis:1980hg,CDF:2001fdy,CMS:2013qur} cross sections at midrapidity as a function of collision energy in \pp\ collisions. Vertical bars around data points represent total uncertainties.}
    \label{fig:rhic_qq_xsec_pp}
\end{figure}

Differential cross sections of heavy quarkonia in \pp\ collisions have been extensively measured at RHIC. The left and right panels of Fig.~\ref{fig:rhic_qq_xsec_pp} present the inclusive \jpsi\ and \ups\ cross sections, where \ups\ denotes $\Upsilon$(1S), $\Upsilon$(2S), and $\Upsilon$(3S) combined, at midrapidity ($y=0$) in \pp\ collisions at $\sqrt{s}=200$ and 510 GeV~\cite{PHENIX:2020dqu,PHENIX:2006aub,STAR:2019vkt,STAR:2021zvb,STAR:2013kwk,STAR:2025ywy,PHENIX:2014tbe}, together with world data~\cite{CDF:2004jtw,ALICE:2021dtt,ALICE:2019pid,ALICE:2012vup,ALICE:2011vrm,CCOR:1979gxw,Kourkoumelis:1980hg,CDF:2001fdy,CMS:2013qur}.
As expected, a clear increase in the heavy quarkonium production cross section with collision energy is observed. For the \ups\ states, a NLO calculation with the Color Evaporation Model (CEM)~\cite{Frawley:2008kk}, as shown in the right panel, describes the data reasonably well over the covered energy range.

\begin{figure}
    \centering
    \includegraphics[width=0.45\linewidth]{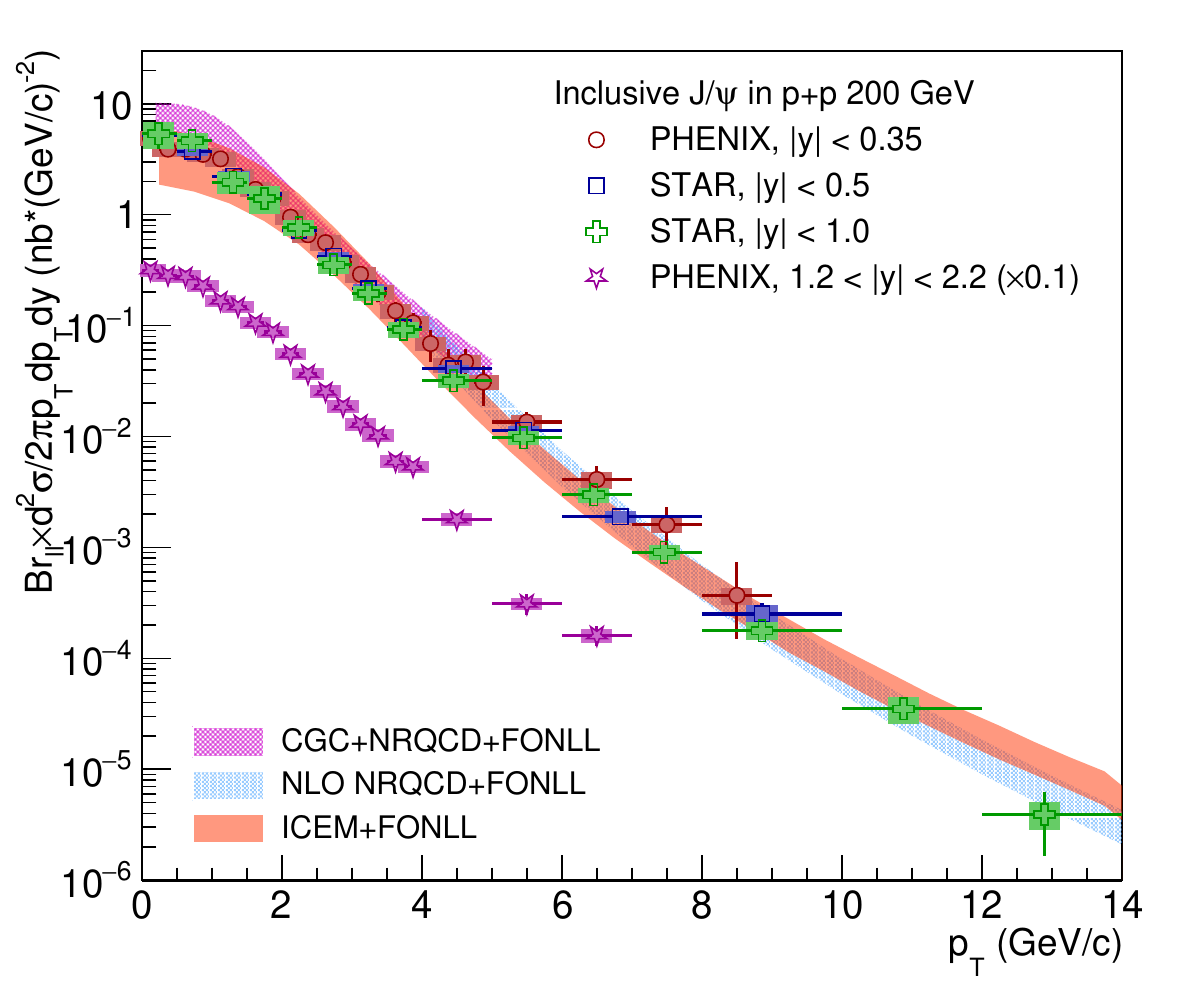}
    \includegraphics[width=0.45\linewidth]{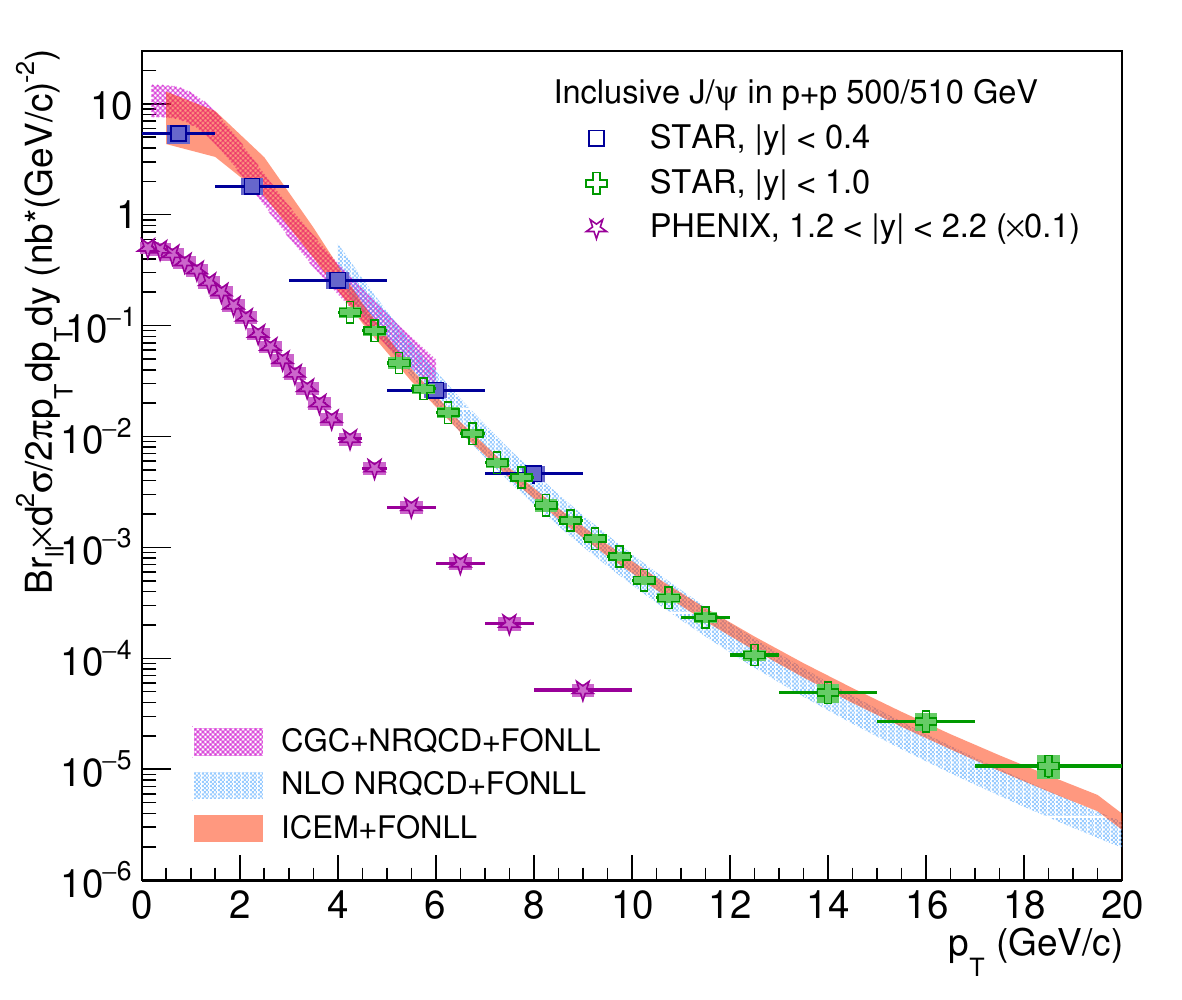}
    \caption{Inclusive \jpsi\ cross section within various rapidity ranges as a function of \jpsi\ \pt\ in 200 (left) and 500/510 GeV (right) \pp\ collisions~\cite{PHENIX:2020dqu,PHENIX:2006aub,STAR:2019vkt,STAR:2021zvb,PHENIX:2019ihw,PHENIX:2011gyb,STAR:2018smh}. Global uncertainties on the order of 10\%, but vary slightly between different measurements, are not shown. Theoretical calculations from CGC+NRQCD~\cite{Ma:2014mri}, NLO NRQCD~\cite{Ma:2010jj,Ma:2010yw} as well as ICEM~\cite{Ma:2016exq} are shown for comparison, where feed-down contributions from bottom-hadron decays are added using FONLL calculations~\cite{Cacciari:1998it,Cacciari:2001td}.}
    \label{fig:rhic_jpsi_xsec_vs_pt_pp}
\end{figure}

\begin{figure}[htb]
	\centering 
	\includegraphics[width=0.5\textwidth]{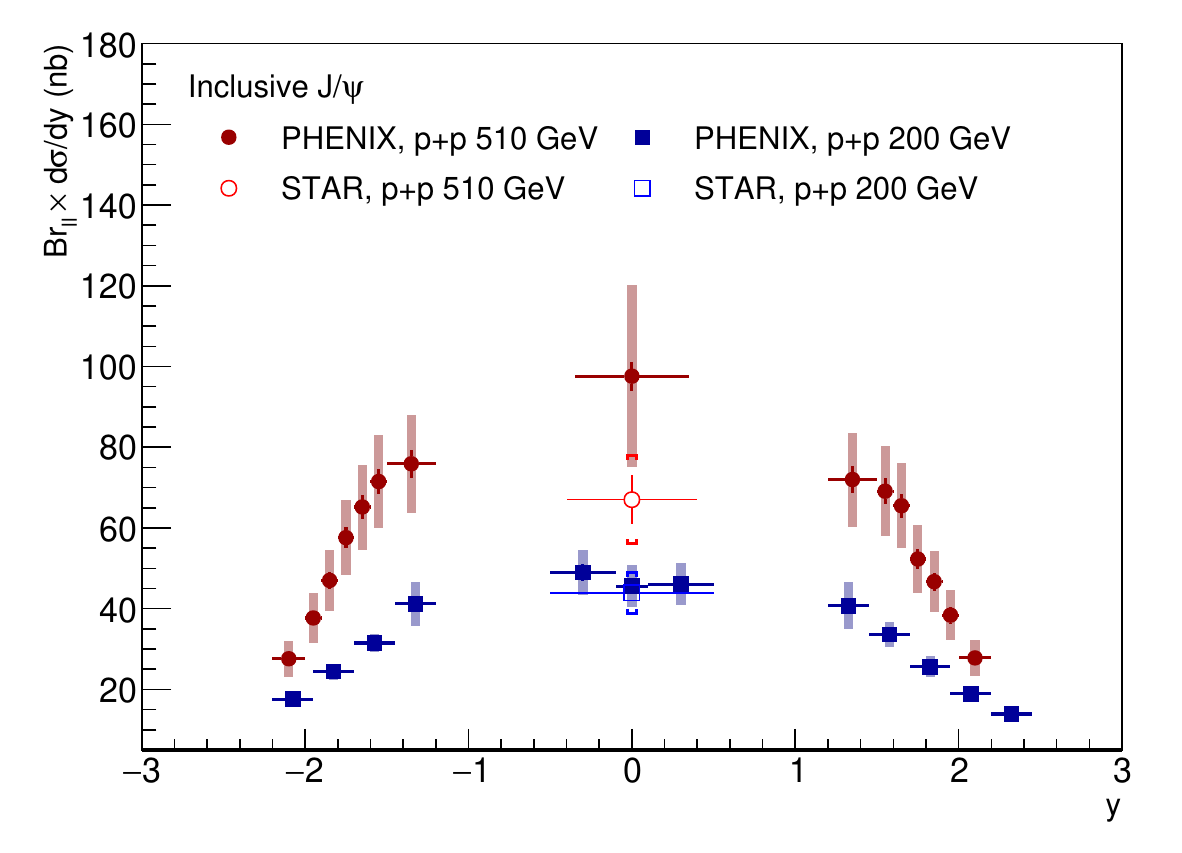}	
	\caption{Rapidity distribution of inclusive \jpsi\ in 200 and 510 GeV \pp\ collisions~\cite{PHENIX:2020dqu,STAR:2019vkt,STAR:2021zvb,PHENIX:2019ihw,PHENIX:2011gyb}. Both filled boxes and brackets around data points show systematic uncertainties. Horizontal bars indicate rapidity intervals over which data are measured. A global 8\% and 10\% uncertainty is not shown for STAR and PHENIX results, respectively.}
	\label{fig:rhic_jpsi_rap_pp}%
\end{figure}

The double-differential cross section of inclusive \jpsi\ production as a function of \jpsi\ \pt\ is shown in Fig.~\ref{fig:rhic_jpsi_xsec_vs_pt_pp}~\cite{PHENIX:2020dqu,PHENIX:2006aub,STAR:2019vkt,STAR:2021zvb,PHENIX:2019ihw,PHENIX:2011gyb,STAR:2018smh}. The left and right panels correspond to \pp\ collisions at $\sqrt{s}=200$ and 500/510 GeV, respectively. Within each panel, different markers represent measurements at different \jpsi\ rapidities, reflecting the different acceptances of the STAR and PHENIX detectors. The measurements at midrapidity ($|y|<1$) are consistent with each other despite slight differences in kinematic coverage. Theoretical calculations based on the non-relativistic QCD (NRQCD) framework are performed at NLO~\cite{Ma:2010jj,Ma:2010yw} for high-\pt\ prompt \jpsi, while the NRQCD framework is coupled with the Color Glass Condensate (CGC) formalism to predict the \jpsi\ cross section at low \pt~\cite{Ma:2014mri}. The non-prompt contribution is calculated with FONLL~\cite{Cacciari:1998it,Cacciari:2001td} and added to the NRQCD results. The calculations are seen to agree with the data within uncertainties. Similar calculations using the improved CEM (ICEM)~\cite{Ma:2016exq} also provide a reasonable description of the measured cross section after the addition of the FONLL contribution. On the other hand, the cross sections at forward rapidity ($1.2<|y|<2.2$) are systematically lower, indicating a decrease in the production cross section toward larger rapidities. 
This trend is further corroborated by the inclusive \jpsi\ cross section, integrated over \pt, measured as a function of rapidity~\cite{PHENIX:2019ihw,PHENIX:2011gyb}, as shown in Fig.~\ref{fig:rhic_jpsi_rap_pp}.

\begin{figure}[htb]
	\centering 
	\includegraphics[width=0.95\textwidth]{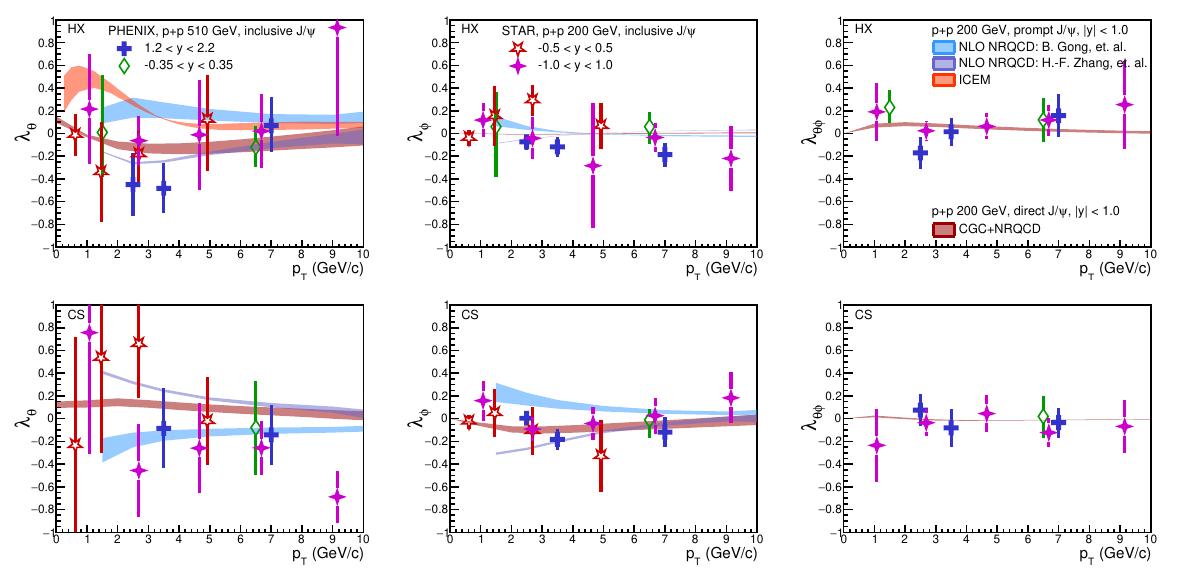}	
	\caption{Inclusive \jpsi\ polarization parameters, $\lambda_{\theta}$, $\lambda_{\phi}$ and $\lambda_{\theta\phi}$, in Helicity (top) and Collins-Soper (bottom) frames as measured in \pp\ collisions~\cite{STAR:2020igu,PHENIX:2020dqu,PHENIX:2016rps}. Vertical bars around data points represent total uncertainties. Calculations from different theoretical approaches are shown for comparison: NLO NRQCD~\cite{Gong:2012ug,Zhang:2014ybe}, CGC+NRQCD~\cite{Ma:2018qvc} and ICEM~\cite{Cheung:2018tvq}.}
	\label{fig:rhic_jpsi_pol}%
\end{figure}

Another powerful probe of the quarkonium production mechanism is through its polarization~\cite{Faccioli:2010kd}. Figure~\ref{fig:rhic_jpsi_pol} compiles measurements of the polarization parameters ($\lambda_{\theta}$, $\lambda_{\phi}$, and $\lambda_{\theta\phi}$) for inclusive \jpsi\ as a function of \pt\ in both the Helicity and Collins–Soper frames~\cite{PHENIX:2020dqu,STAR:2020igu,PHENIX:2016rps}. The measured polarization parameters are consistent with zero within uncertainties at both mid- and forward rapidities, and in \pp\ collisions at $\sqrt{s}=200$ and 510 GeV. Theoretical predictions based on NLO NRQCD~\cite{Gong:2012ug,Zhang:2014ybe}, CGC+NRQCD~\cite{Ma:2018qvc} and ICEM~\cite{Cheung:2018tvq} agree with data within uncertainties. 

Furthermore, the dependence of the inclusive \jpsi\ production on event activity is measured in \pp\ collisions at \sqrts\ = 200 GeV~\cite{STAR:2018smh,PHENIX:2024dqs}. Comparisons with PYTHIA predictions~\cite{Sjostrand:2007gs,Aguilar:2021sfa} suggest that multi-parton interactions are needed to describe the experimental data with PYTHIA. 

\subsection{Small system collisions}
Similar to open heavy flavor, CNM effects~\cite{Arleo:2025oos} also affect quarkonium production in heavy-ion collisions. Typical CNM effects include nPDF modifications~\cite{Klasen:2023uqj}, partonic energy loss in cold nuclear matter~\cite{Arleo:2013zua}, and the breakup (or absorption) of quarkonia through interactions with nucleons in the colliding nuclei or with co-moving final-state hadrons~\cite{Ferreiro:2011xy,Ferreiro:2014bia}. 

\begin{figure}[htb]
	\begin{minipage}{0.49\linewidth}
	    \includegraphics[width=1\textwidth]{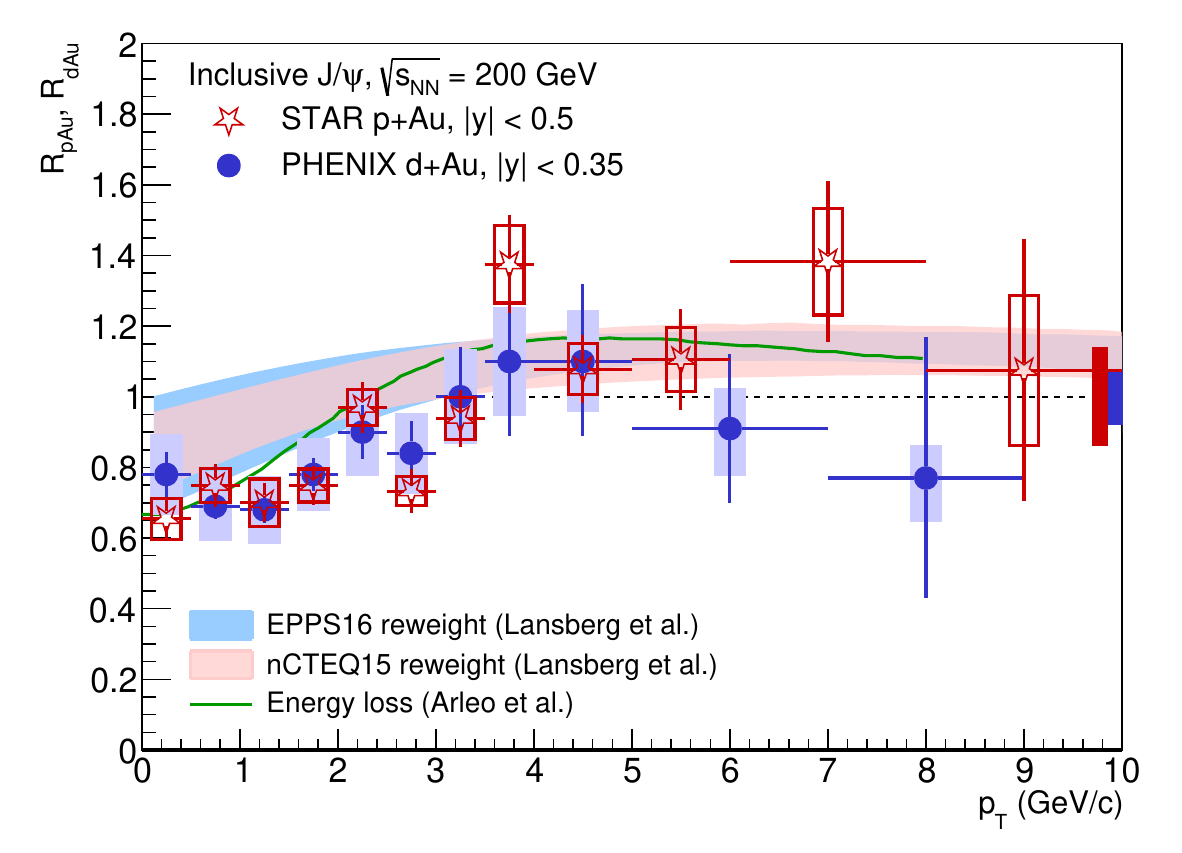}
	\end{minipage}
        \begin{minipage}{0.49\linewidth}
            \includegraphics[width=1\linewidth]{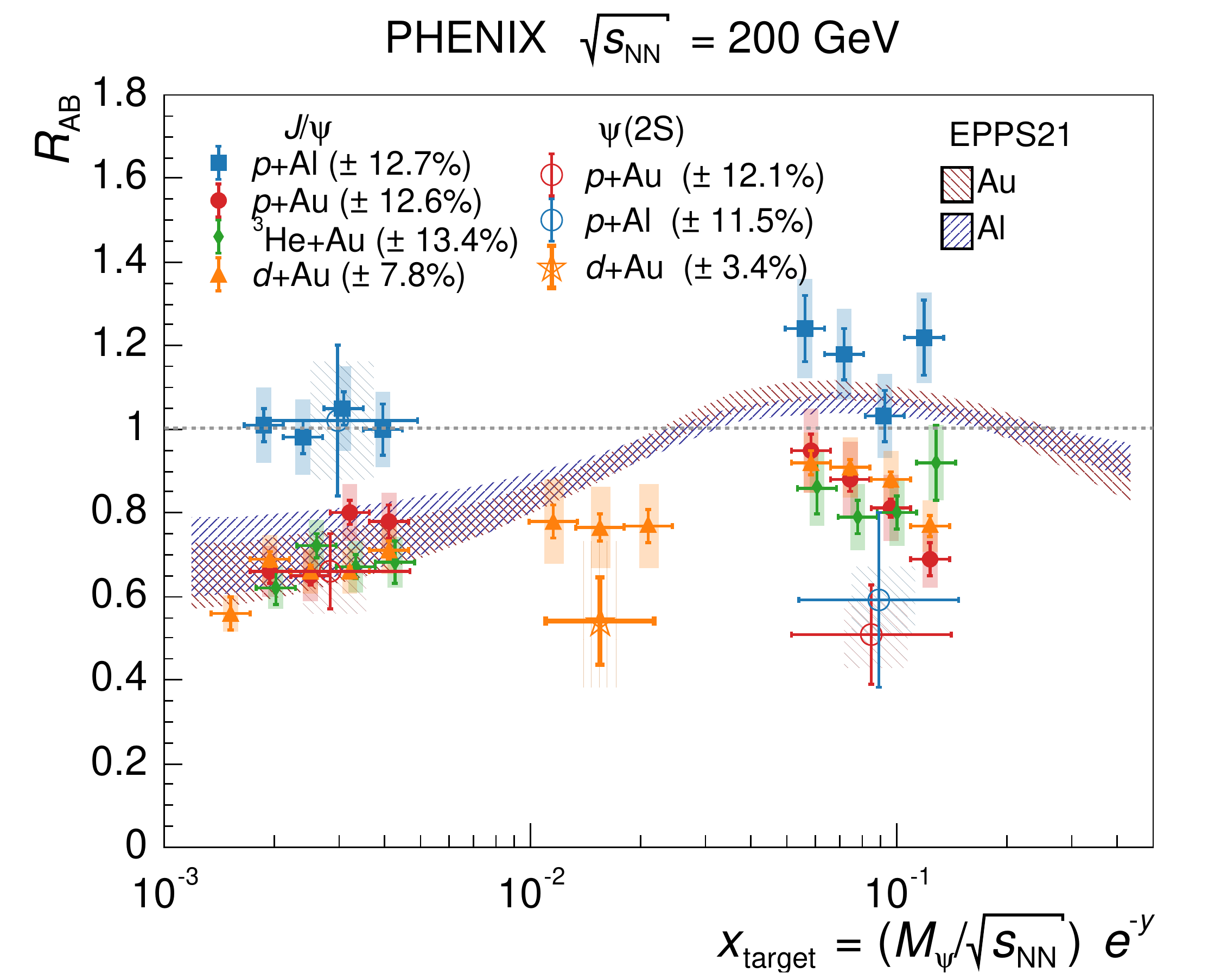}
        \end{minipage}
	\caption{Left: inclusive \jpsi\ \rpa\ measured in \pAu~\cite{STAR:2021zvb} and \dAu~\cite{PHENIX:2012czk} collisions at \sqrtsnn\ = 200 GeV at midrapidity. Vertical bands at unity display global uncertainties. Model calculations incorporating nPDF effects~\cite{Kusina:2017gkz} and energy loss~\cite{Arleo:2013zua} are also shown for comparison. Right: $x_{\rm target}$ dependence of the \jpsi and $\psi(2S)$ nuclear modification factors in various small-system collisions~\cite{PHENIX:2010hmo,PHENIX:2019brm, PHENIX:2013pmn, PHENIX:2022nrm} along with the EPPS21 nPDF predictions~\cite{Eskola:2021nhw}. Global uncertainties are indicated in the legends.}
\label{fig:rhic_jpsi_RpA}%
\end{figure}

Figure~\ref{fig:rhic_jpsi_RpA}, left panel, shows the inclusive \jpsi\ \rpa\ at midrapidity measured at RHIC in \pAu~\cite{STAR:2021zvb} and \dAu~\cite{PHENIX:2012czk} collisions at \sqrtsnn\ = 200 GeV. A suppression of approximately 30\% is observed at $p_{\rm T} < 2$ \gevc, which gradually diminishes with increasing \pt. Above $p_{\rm T} \sim 3$ \gev, \rpa\ becomes consistent with unity within uncertainties, suggesting that CNM effects are almost absent in this kinematic region. Theoretical calculations incorporating nPDF modifications~\cite{Kusina:2017gkz} or partonic energy loss in cold nuclear matter~\cite{Arleo:2013zua} provide a reasonable description of the data. However, both calculations tend to lie slightly below the experimental data for $p_{\rm T} < 3$ \gev, indicating that multiple CNM mechanisms may contribute simultaneously.

The rapidity dependence of the \jpsi\ nuclear modification factor ($R_{\rm AB}$) is also measured at RHIC in $p$+Al, \pAu, \dAu, and $^{3}$He+Au collisions~\cite{PHENIX:2010hmo,PHENIX:2019brm}, probing an extended range of Bjorken-$x$ carried by gluons in the target nucleus ($x_{\rm target}$). Assuming that quarkonia production is dominated by the leading-order process $gg \rightarrow Q\bar{Q}$, $x_{\rm target}$ can be approximated as
\begin{equation}
x_{\rm target} = \frac{M_{\psi}}{\sqrt{s_{\rm NN}}} e^{-y},
\end{equation}
where $M_{\psi}$ is the \jpsi\ or $\psi(2S)$ mass and $y$ is its rapidity. Figure~\ref{fig:rhic_jpsi_RpA} (right) shows the \jpsi\ $R_{\rm AB}$ as a function of $x_{\rm target}$. The measurements exhibit clear evidence of small-$x$ suppression and a gradual reduction of the suppression for $x_{\rm target} > 10^{-2}$. Comparisons with calculations based on the EPPS21 nPDF set~\cite{Eskola:2021nhw}, which incorporates constraints from LHC $p$+Pb data, indicate that nPDF effects alone can describe the data at $x_{\rm target} \lesssim 10^{-2}$ for the Au target. At larger $x_{\rm target}$, the observed suppression in Au target exceeds the EPPS21 nPDF expectation and differs from the enhancement observed for leptons from heavy-flavor decays at mid- and backward rapidities, as shown in Fig.~\ref{fig:HF_RdA}. In contrast, nuclear effects in $p$+Al collisions are consistent with zero within uncertainties, highlighting a clear nuclear dependence of CNM effects, which is not reproduced by the EPPS21 nPDF predictions.

Nuclear modifications of the excited charmonium states, $\psi$(2S) and $\chi_{c}$, have also been measured in 200 GeV \pau\ and \dAu\ collisions at RHIC~\cite{PHENIX:2013pmn,PHENIX:2022nrm}. $\psi$(2S) measurements are shown in the right panel of Fig.~\ref{fig:rhic_jpsi_RpA} as open symbols. At forward rapidity ($1.2<y<2.2$) or small $x_{\rm target}$, the $\psi$(2S) \rpa\ is consistent with that of the \jpsi, supporting the expectation that initial-state effects are similar for the two states. Toward backward rapidity, however, $\psi$(2S) exhibits increasingly stronger suppression than the \jpsi, suggesting enhanced final-state nuclear effects, such as nuclear absorption or co-mover breakup, for the more weakly bound $\psi$(2S) state. The uncertainty on the $\chi_{c}$ measurement is currently too large to draw definitive conclusions~\cite{PHENIX:2013pmn}.

\subsection{Heavy-ion collisions}
\subsubsection{\jpsi\ nuclear modification factor}
\begin{figure}
    \centering
    \includegraphics[width=0.5\linewidth]{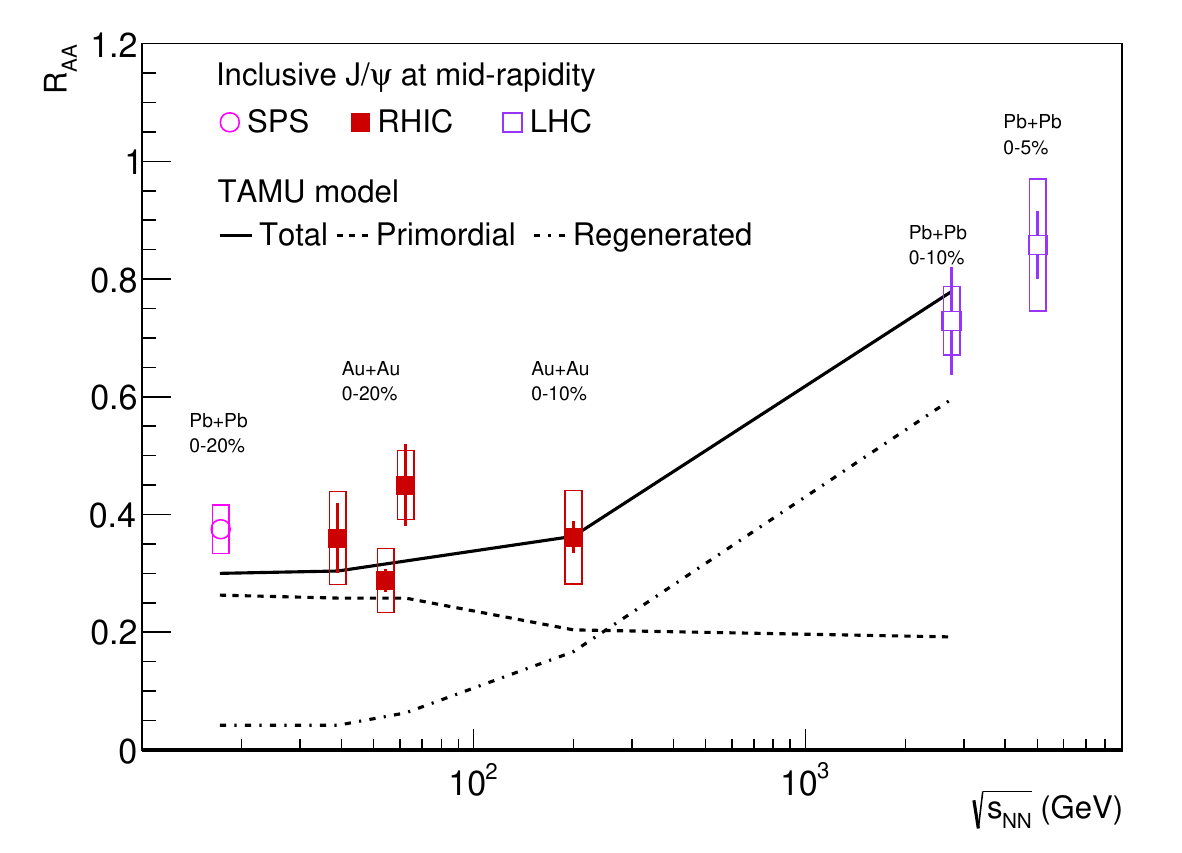}
    \caption{Inclusive \jpsi\ \raa\ as a function of collision energy in central collisions ~\cite{NA50:2000brc,Kluberg:2005yh,STAR:2016utm,STAR:2019fge,ALICE:2013osk,ALICE:2023gco}. Transport model calculations from the TAMU group~\cite{Zhao:2010nk} are compared to data.}
    \label{fig:rhic_jpsi_RAA_energy}
\end{figure}

\begin{figure}
    \centering
    \includegraphics[width=0.45\linewidth]{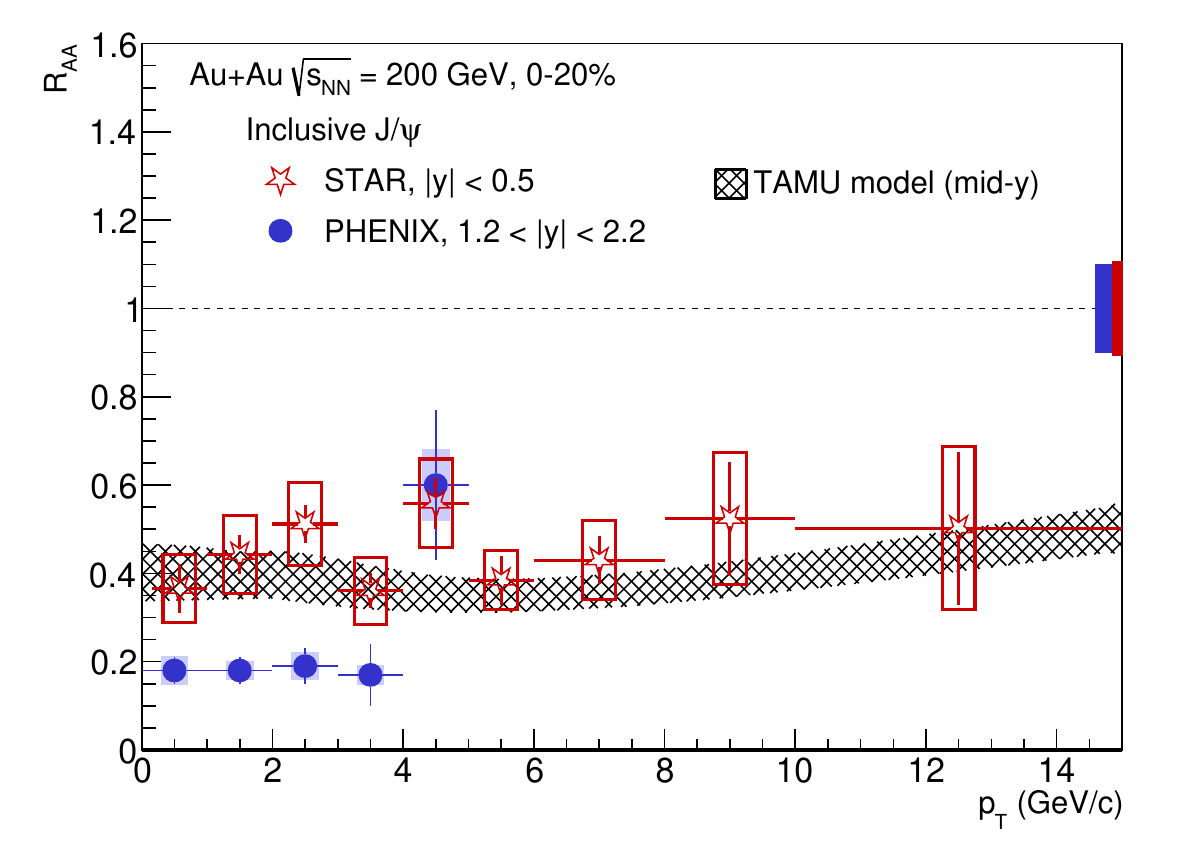}
    \includegraphics[width=0.45\linewidth]{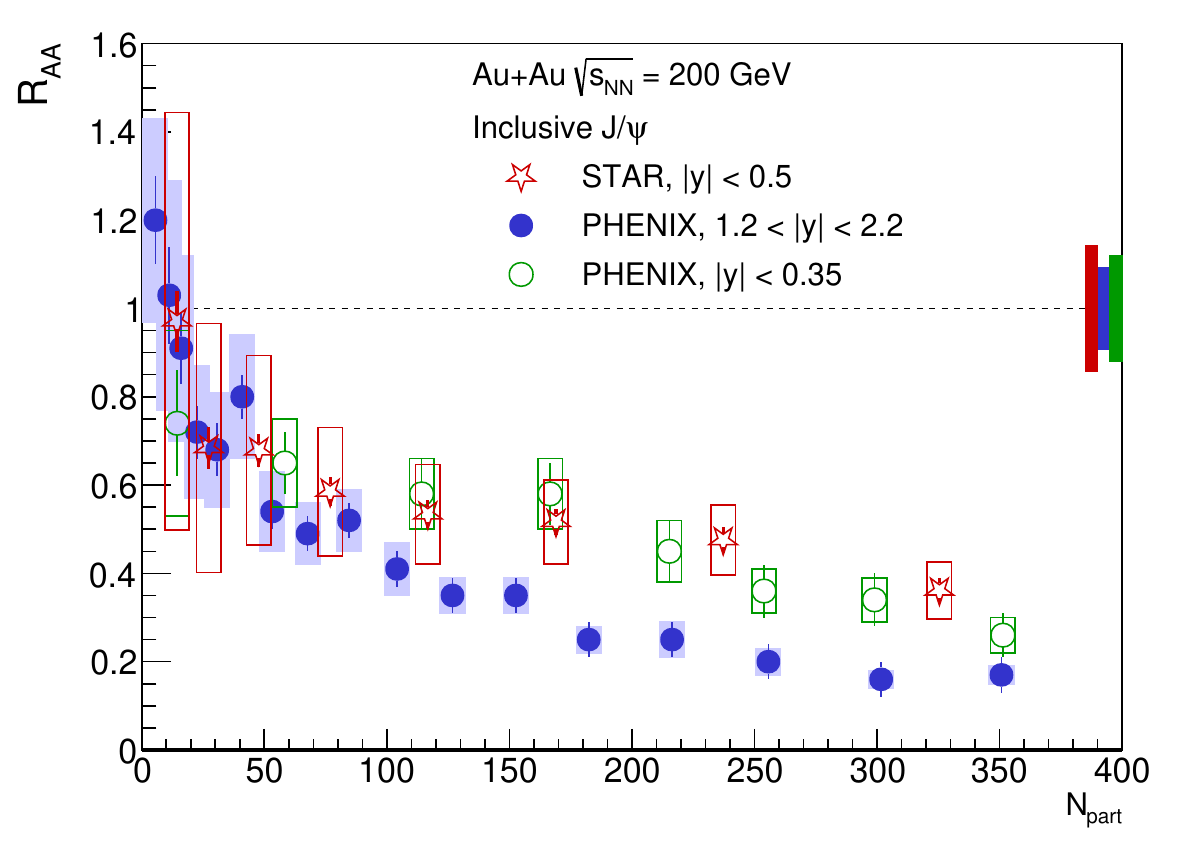}
    \caption{Inclusive \jpsi\ \raa\ as a function of \pt\  (left) and \npart\ (right) at mid and forward rapidities in 200 GeV \AuAu\ collisions~\cite{STAR:2019fge, PHENIX:2006gsi, PHENIX:2011img}.}
    \label{fig:rhic_jpsi_RAA}
\end{figure}

Figure~\ref{fig:rhic_jpsi_RAA_energy} presents the inclusive \jpsi\ \raa\ in central heavy-ion collisions as a function of collision energy, measured by experiments at the SPS, RHIC, and the LHC~\cite{NA50:2000brc,Kluberg:2005yh,STAR:2016utm,STAR:2019fge,ALICE:2013osk,ALICE:2023gco}. For \sqrtsnn\ $\le$ 200 GeV, the \jpsi\ \raa\ remains approximately constant. This behavior reflects the interplay between several competing mechanisms: CNM effects, which tend to decrease with increasing collision energy, and in-medium dissociation and regeneration, both of which become stronger at higher temperatures. In contrast, a pronounced increase of \raa\ is observed from 200 GeV to 5.02 TeV. At these higher energies, the charm quark production cross section rises substantially, leading to a significant contribution from regeneration of \jpsi\ in the deconfined medium and driving the observed enhancement of \raa\ at the LHC. A transport model calculation by the TAMU group~\cite{Zhao:2010nk}, which incorporates CNM effects, in-medium dissociation, and regeneration within a unified framework, is also shown in Fig.~\ref{fig:rhic_jpsi_RAA}. The regeneration contribution is displayed separately. The model provides a reasonable description of the overall collision-energy dependence, supporting the interpretation based on the interplay between CNM effects, dissociation and regeneration mechanisms.

Differential measurements of inclusive \jpsi\ \raa\ as a function of \pt\ at mid and forward rapidities are shown in the left panel of Fig.~\ref{fig:rhic_jpsi_RAA} for 0-20\% central \AuAu\ collisions at \sqrtsnn\ = 200 GeV~\cite{STAR:2019fge,PHENIX:2006gsi}. 
At midrapidity, the measured \raa\ is approximately 0.5 over a broad \pt\ range, indicating that the inclusive \jpsi\ yield is suppressed by about a factor of two in central \AuAu\ collisions. Only a mild \pt\ dependence is observed. Part of the suppression at $p_{\rm T} < 3$ \gev\ can be attributed to CNM effects, as illustrated in Fig.~\ref{fig:rhic_jpsi_RpA}. However, at higher \pt, the dominant contribution arises from the in-medium dissociation of \jpsi\ and excited charmonium states. The observed suppression therefore provides direct evidence for charmonium dissociation in the deconfined medium formed in these collisions. Calculations from the TAMU transport model~\cite{Zhao:2010nk} describe the data at midrapidity reasonably well. The model suggests that the slight increase of \raa\ with increasing \pt\ can be attributed to several effects. One is the formation-time effect: higher-\pt\ \jpsi\ have shorter interaction times with the medium and are more likely to escape before being dissociated. In addition, feed-down from bottom hadrons, which are expected to experience weak suppression in the QGP due to the large bottom-quark mass, contributes more significantly at higher \pt, further enhancing \raa\ in this region. On the other hand, the suppression is observed to be stronger at forward rapidity than that at midrapidity, despite the expectation that QGP effects should decrease toward forward rapidity. This ``reversed" rapidity ordering may be attributed to an enhanced regeneration contribution at midrapidity, analogous to the reduced suppression observed at higher collision energies, as illustrated in Fig.~\ref{fig:rhic_jpsi_RAA_energy}. A comparison of \pt-integrated \jpsi\ \raa\ as a function of centrality, characterized by the number of participating nucleons (\npart), at mid and forward rapidities is illustrated in the right panel of Fig. \ref{fig:rhic_jpsi_RAA}~\cite{STAR:2019fge,PHENIX:2006gsi,PHENIX:2011img}. Both measurements exhibit increasing suppression from peripheral to central collisions, consistent with the expectation of stronger QGP effects in more central events. Again, a larger suppression is observed at forward rapidity.

\subsubsection{Sequential suppression}
Since the dissociation of quarkonium states in the QGP depends on the intrinsic properties of the quarkonium states, systematic measurements of the suppression pattern across different quarkonium states thus provide a powerful tool for constraining the properties of the QGP. At the same time, such measurements are essential for a proper interpretation of the ground-state suppression. 

\begin{figure}[htb]
	\centering 
    	\includegraphics[width=0.47\textwidth]{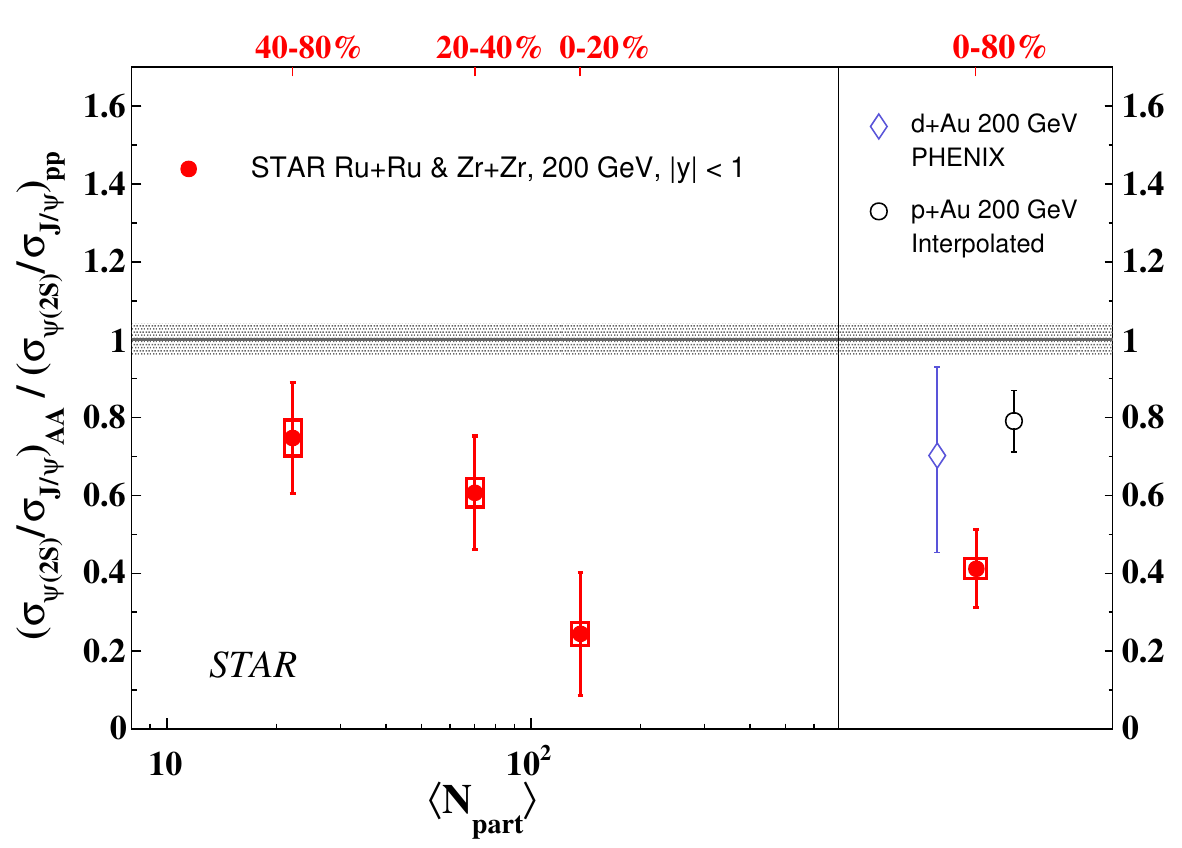}	
	\includegraphics[width=0.45\textwidth]{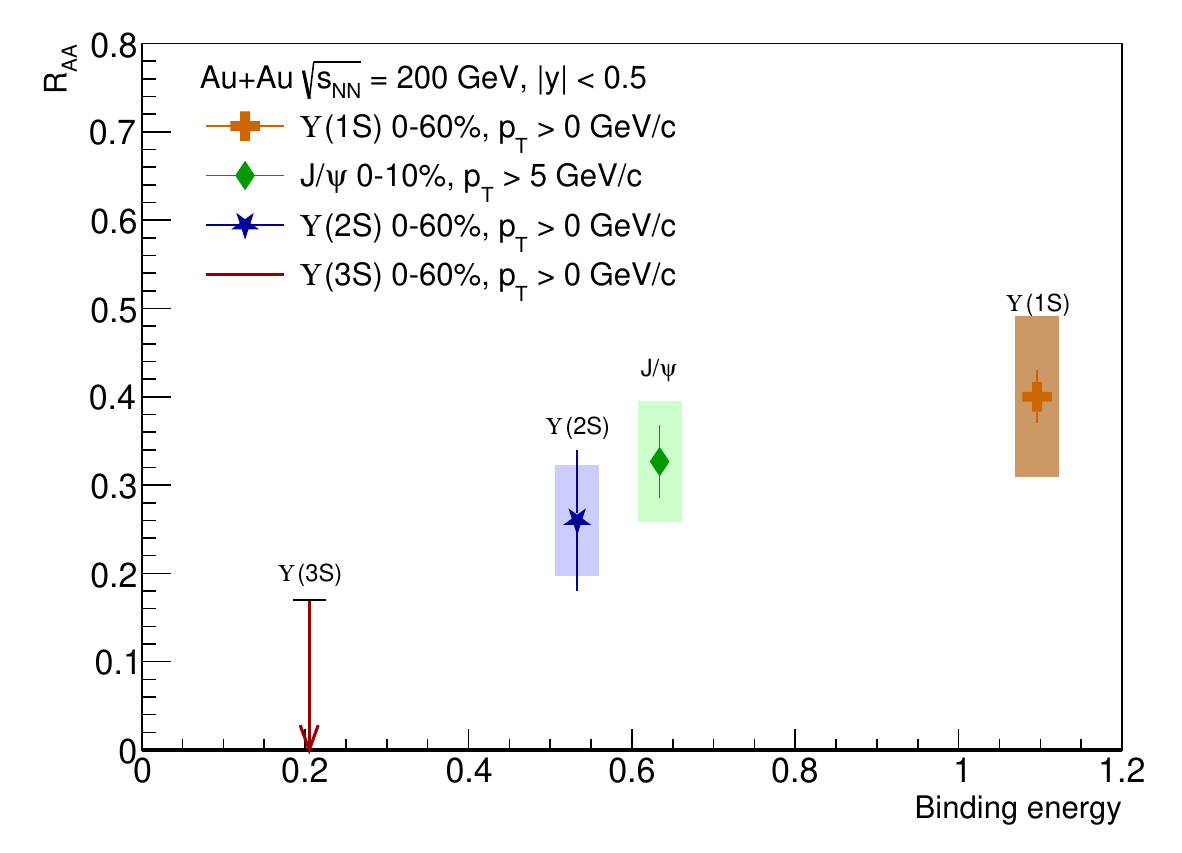}	
	\caption{Left: Ratio of $\psi(2S)$ to \jpsi\ yields in 200 GeV Ru+Ru and Zr+Zr collisions relative to that in \pp\ collisions as a function of collision centrality~\cite{STAR:2025imj}. Corresponding ratios from \dAu\ and \pAu\ collisions, for which total uncertainties are indicated by vertical bars, are shown for comparison. Here, the \pAu\ result is obtained from an interpolation of world data~\cite{STAR:2025imj}. Right: \raa\ of quarkonium states with different binding energies at midrapidity in \AuAu\ collisions at \sqrtsnn\ = 200 GeV~\cite{STAR:2019fge,STAR:2022rpk}. The upper limit for \ups(3S) is shown at the 95\% confidence level.} 
	\label{fig:Quarkonia_RHIC_RAA}%
\end{figure}

Measurement of sequential suppression in the charmonium sector is shown in the left panel of Fig.~\ref{fig:Quarkonia_RHIC_RAA}. The $\psi(2S)$ to \jpsi\ yield ratio within $|y|<1$ in \RuRu\ and \ZrZr\ collisions at \sqrtsnn\ = 200 GeV, relative to that in \pp\ collisions, is presented as a function of \npart~\cite{STAR:2025imj}. This double ratio is found to be below unity, indicating that the more weakly bound $\psi(2S)$ state is more strongly suppressed than the more tightly bound \jpsi\ in these collisions. Furthermore, the decreasing trend of the double ratio from peripheral to central collisions is consistent with the expectation of increasing medium effects in more central events. As in the case of \raa, potential contributions from CNM effects must be considered when extracting QGP properties from the double ratio. A similar double-ratio measurement in \dAu\ collisions is also shown in the left panel, although it is subject to large experimental uncertainties. An interpolated value for 200 GeV \pAu\ collisions~\cite{STAR:2025imj}, derived from world data, is presented with significantly reduced uncertainty compared to the direct \dAu\ measurement. The observation that the double ratio in \RuRu\ and \ZrZr\ collisions is approximately a factor of two smaller than the interpolated \pAu\ result suggests the presence of additional QGP-induced sequential suppression beyond what can be attributed to CNM effects alone. 

The sequential suppression in the bottominum section is likewise measured at RHIC. Figure~\ref{fig:Quarkonia_RHIC_RAA}, right panel, shows \pt-integrated \raa\ at midrapidity for $\Upsilon$(1S), $\Upsilon$(2S), and $\Upsilon$(3S), in 0-60\% \AuAu\ collisions at \sqrtsnn\ = 200 GeV~\cite{STAR:2022rpk}. Although the experimental uncertainties are sizable, a trend of decreasing \raa\ with decreasing binding energy is observed. For comparison, the \raa\ of high-\pt\ \jpsi\ is also shown in the figure. Its suppression level follows the overall binding-energy hierarchy, further reinforcing the sequential suppression picture across both the charmonium and bottomonium families.

\subsubsection{Other developments} In addition to \raa, inclusive \jpsi\ $v_{2}$ has been measured at both mid- and forward rapidities in 200 GeV \AuAu\ collisions~\cite{STAR:2012jzy,PHENIX:2024axj}. Within the current experimental uncertainties, the measured $v_{2}$ values are consistent with zero over the explored \pt\ range, suggesting that \jpsi\ production is not dominated by the coalescence of thermalized charm quarks at  high \pt.

The inclusive \jpsi\ polarization parameters have also been measured in 200 GeV \RuRu\ and \ZrZr\ collisions~\cite{STAR:2026puo}. The polarization parameters are consistent with zero and agree, within uncertainties, with those measured in \pp\ collisions (Fig.~\ref{fig:rhic_jpsi_pol})~\cite{STAR:2020igu}, indicating no significant modification of \jpsi\ polarization in the nuclear medium.

\subsection{Summary}
\begin{itemize}
    \item  In $p$+$p$ collisions, quarkonium production cross sections and polarization at RHIC can be qualitatively described by NRQCD and CEM calculations within current experimental and theoretical uncertainties. These results provide a solid baseline for understanding quarkonium behavior in the hot QCD medium.

    \item Cold nuclear matter effects are seen to significantly influence quarkonium production, particularly at low \pt. In addition, excited charmonium states exhibit stronger suppression than the corresponding ground states, an effect that must be properly accounted for when interpreting heavy-ion collision results.

    \item In heavy-ion collisions, multi-differential measurements of inclusive \jpsi\ suppression as functions of collision energy, event centrality, \jpsi\ \pt, and rapidity indicate that both cold and hot nuclear matter effects are required to describe the observed patterns. The latter include quarkonium dissociation in the QGP as well as regeneration from deconfined heavy quarks. Furthermore, a clear sequential suppression pattern is observed for both charmonium and bottomonium families, with more weakly bound states experiencing stronger suppression. These measurements provide important constraints on the thermodynamic properties of the QGP and on the modification of the heavy-quark potential under extreme conditions.
\end{itemize}

\section{Future Perspective and Connection to EIC}

\subsection{Heavy flavor program at RHIC in the near future}
The final years of \rhic saw several upgrades to the \starexp experiment through the inner TPC (iTPC) which improves the tracking efficiency and momentum resolution, and the Data Acquisition System which enables high throughput data collection. The new \sphenix experiment, designed for precision heavy-flavor and jet measurements at midrapidity~\cite{Campbell:2016cea}, was commissioned and completed physics data taken in \pp, \OO~and \AuAu collisions. Although RHIC operations have concluded, the large data sets collected by \starexp and \sphenix will continue to provide important insights into heavy-flavor physics for years to come.

A unique feature of the \sphenix\ experiment is its streaming readout architecture, in which the tracking detectors record all data arriving in the data acquisition system independently of the hardware trigger. This capability is particularly important for open heavy-flavor measurements at RHIC energies, where the production spectra of charm and bottom hadrons peak at relatively low \pt. By avoiding trigger thresholds that preferentially reject low-\pt\ particles, streaming readout substantially improves the statistical reach for open heavy-flavor measurements. The streaming readout was successfully deployed during the RHIC \pp and \OO\ runs, while it was not required for the \AuAu\ run because the collision rate was comparable to the trigger bandwidth.

\begin{figure}
    \centering
    \includegraphics[width=0.45\linewidth]{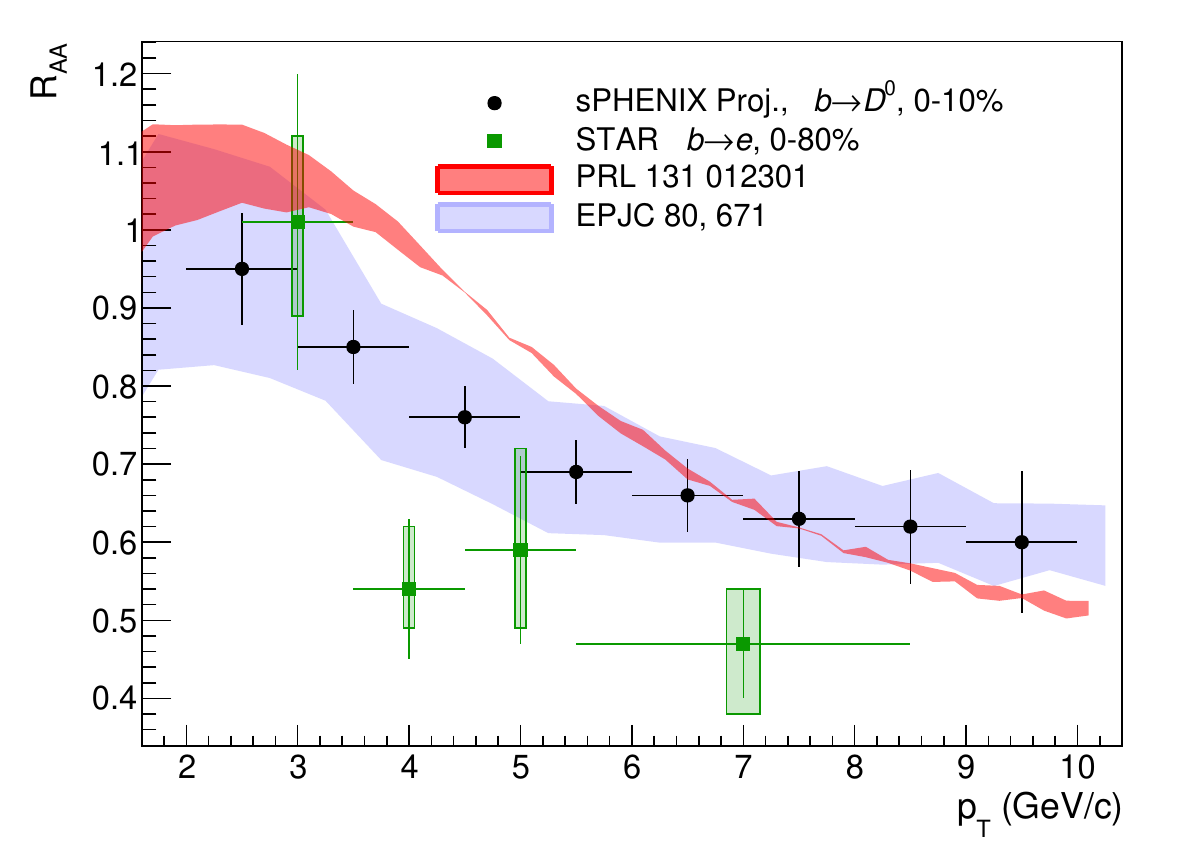}
    \includegraphics[width=0.45\linewidth]{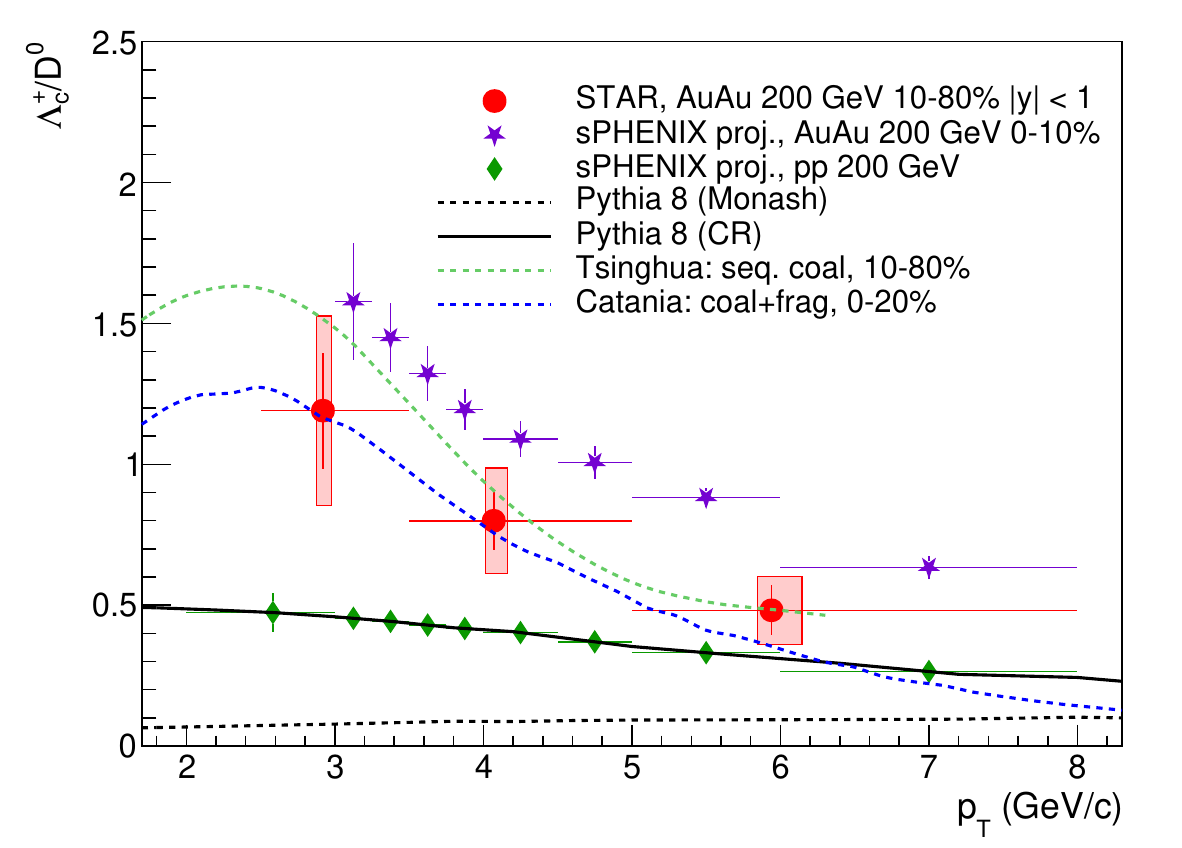}
    \caption{Left: Projected statistical precision of \sphenix\ on \RAA\ of non-prompt $D^0$ mesons from bottom-hadron decays, compared with previous measurements of bottom-hadron decayed electrons from the \starexp\ collaboration. Right: Projected statistical sensitivity of \sphenix\ to the $\Lambda_c^+/D^0$ production ratio in \pp\ and \AuAu\ collisions. Various model calculations are also shown for comparison.}
    \label{fig:sphenix_open_HF_projections}
\end{figure}

\begin{figure}
    \centering
    \includegraphics[width=0.49\linewidth]{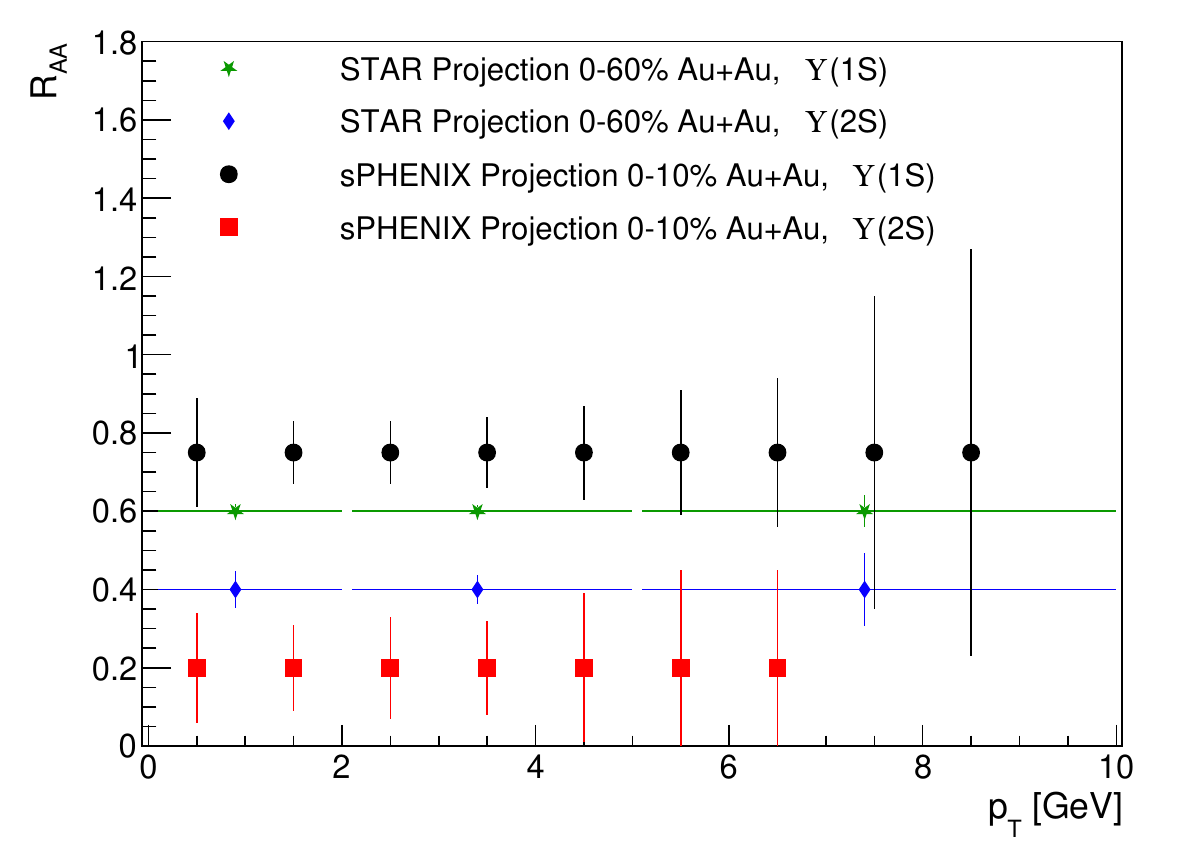}
    \includegraphics[width=0.49\linewidth]{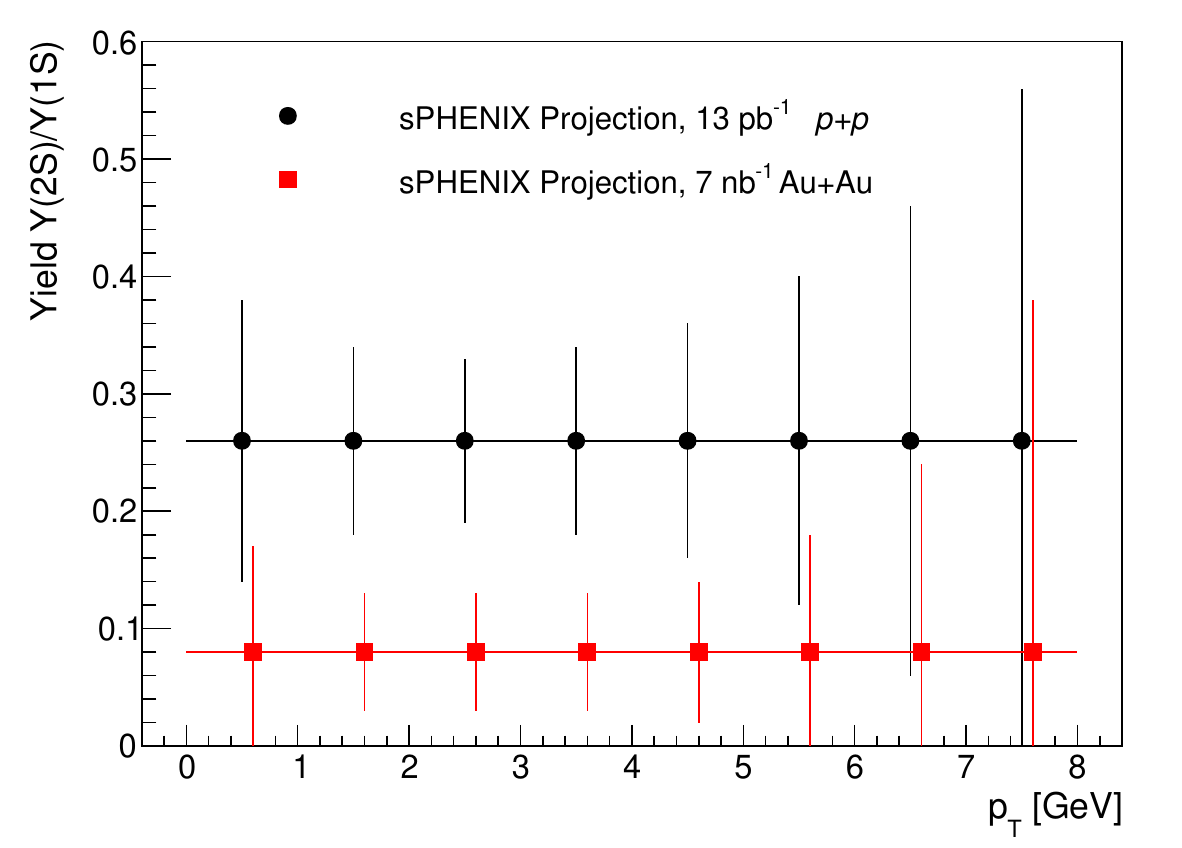}
    \caption{Left: Projected statistical sensitivities of STAR and \sphenix\ to \RAA\ of the $\Upsilon$(1S) and $\Upsilon$(2S) states. Right: Projected statistical sensitivity of \sphenix\ to the $\Upsilon(\rm 2S)/\Upsilon(1S)$ production ratio in \pp\ and \AuAu\ collisions.}

    \label{fig:sphenix_upsilon_ratio}
\end{figure}

The remaining RHIC heavy-flavor program will focus on several outstanding physics questions that were previously limited by statistical precision:
\begin{itemize}
\item \textbf{Bottom quark transport in the QGP:} Precision measurements of open bottom (tagged via non-prompt $D$, $J/\psi$ or $e$ etc.) \RAA\ and $v_2$ to constrain the bottom quark diffusion coefficient and the relative importance of collisional and radiative energy losses. The left panel of Fig.~\ref{fig:sphenix_open_HF_projections} shows the projected \sphenix sensitivity to \raa\ of non-prompt $D^0$ hadrons in 0-10\% Au+Au collisions. Previous measurements at RHIC relied on non-prompt electrons to tag bottom-hadron decays (Fig.~\ref{fig:rhic_raa}), whereas \sphenix will reconstruct non-prompt $D^0$ mesons. For illustration, predictions from models that emphasize radiative energy loss~\cite{Li:2019lex} or enhanced feed-down contributions~\cite{He:2019vgs} for LHC energies are also shown. 
\item \textbf{Heavy quark hadronization:} Measurements of charm-baryon and charm-meson production, particularly the $\Lambda_c^+/D^0$ ratios in both \pp and \AuAu\ collisions, to distinguish between fragmentation, coalescence, and color-reconnection mechanisms. The high-statistics \pp\ data collected with the \sphenix streaming-readout system will enable a high-precision measurement of the $\Lambda_c^+/D^0$ ratio, as illustrated in the right panel of Fig.~\ref{fig:sphenix_open_HF_projections}. This measurement will establish the first data-driven \pp\ baseline at RHIC for quantifying medium-induced modifications to charm hadronization. Furthermore, \sphenix is expected to measure the $\Lambda_c^+/D^0$ ratio in \AuAu\ collisions with high precision over a broad kinematic range, as illustrated in Fig.~\ref{fig:sphenix_open_HF_projections}. 
\item \textbf{Sequential quarkonium suppression:} High-statistics measurements of the $\Upsilon$(1S), $\Upsilon$(2S), and $\Upsilon$(3S) states, together with charmonium states including $\psi$(2S), to quantify sequential melting and constrain medium-induced dissociation in the QGP. The projected STAR and \sphenix sensitivities to \raa\ the $\Upsilon$(1S) and $\Upsilon$(2S) states, as well as the $\Upsilon(\rm 2S)/\Upsilon(1S)$ production ratio, are shown in Fig.~\ref{fig:sphenix_upsilon_ratio}. Significant improvements in the measurement of the $\Upsilon$(3S) state, together with the first measurement of $\psi(2S)$ production, in \AuAu\ collisions at RHIC are also anticipated.
\item \textbf{Charmonium production dynamics:} Precision measurements of inclusive \jpsi\ flow, polarization, spin alignment, and differential \raa\ to further constrain the interplay between dissociation and regeneration. Improved measurements of \jpsi\ $v_2$ will further quantify the degree of charm-quark thermalization and the contribution from regeneration, while polarization measurements will probe possible modifications of the \jpsi\ production mechanism in the nuclear medium. Together with the first measurements of $\psi(2S)$ production in \AuAu\ collisions at RHIC, these observables will establish a comprehensive picture of charmonium production and transport in the QGP.

\item \textbf{Spin-dependent heavy-flavor observables:} Measurements of transverse single-spin asymmetries in heavy-flavor production to probe gluon spin-orbit correlations and multi-gluon correlation functions. RHIC is uniquely capable of such measurements through collisions of transversely polarized proton beams. PHENIX reported evidence for a sizable \jpsi\ transverse single-spin asymmetries (TSSAs)~\cite{PHENIX:2010hqq}. Since gluon--quark correlations are expected to generate only small asymmetries, a sizable TSSA would point to significant trigluon correlations~\cite{Kang:2008ih}. The large data set collected by \sphenix\ will enable precision measurements of the $D^0$ TSSA, providing a stringent test of the role of trigluon correlations in charm production.

\item \textbf{Multi-parton interactions:} The observation of significant effects of multi-parton interactions (MPI) on \jpsi\ production \cite{STAR:2018smh,PHENIX:2024dqs} opens an opportunity to study how particle production are affected by MPI and if partons inside the proton and nucleus are correlated. Further analysis of existing data from STAR, PHENIX and \sphenix will enable more detailed studies of quarkonia and heavy-flavor correlations with underlying event activity.
\end{itemize}

\subsection{Connections to EIC}
The heavy-flavor program at RHIC naturally connects to the future Electron-Ion Collider (EIC), where open heavy-flavor production in \ep\ and $e$+A collisions proceeds predominantly through photon--gluon fusion. Unlike heavy-ion collisions, the absence of a hot QCD medium at EIC provides a clean environment in which heavy-quark production, fragmentation, and hadronization can be studied with controlled systematics. Consequently, the EIC will help establish precise baselines for interpreting heavy-flavor measurements at RHIC and disentangle cold nuclear matter effects from QGP-induced modifications.

The EIC will also significantly advance our understanding of gluon dynamics in nucleons and nuclei through precision measurements of open heavy flavor and quarkonia~\cite{AbdulKhalek:2021gbh}. Heavy-flavor production is directly sensitive to the gluon distributions over a broad range of Bjorken-$x$, enabling stringent constraints on nPDFs and gluon saturation at small $x$, and to heavy-quark hadronization in cold nuclear matter~\cite{Li:2020zbk,Kelsey:2021gpk,Boer:2024ylx}. Measurements of charm-hadron production ratios, such as $D_s^+/D^0$ and $\Lambda_c^+/D^0$, in both \ep\ and $e$+A collisions will provide critical information on the universality of heavy-quark fragmentation and the modification of hadronization in nuclei, thereby complementing the corresponding measurements in \pp\ and \AuAu\ collisions at RHIC.

The polarized electron and proton beams at the EIC will further extend the heavy-flavor spin program initiated at RHIC. In particular, spin-dependent observables in open-heavy-flavor and exclusive quarkonium production will provide new insight into the spin and spatial structure of gluons in the nucleon. Measurements of the longitudinal double-spin asymmetry ($A_{\rm LL}$), together with exclusive charmonium production and other spin-dependent observables, will constrain the gluon helicity distribution and generalized parton distributions (GPDs)~\cite{Goloskokov:2024egn}.

\section{Summary}
RHIC concluded 26 years of operation in 2026, leaving a lasting legacy in the study of strongly interacting QCD matter. Heavy-flavor measurements have played a central role in establishing the properties of the strongly coupled quark-gluon plasma by exploiting the unique sensitivity of charm and bottom quarks to the properties of the medium.

Measurements of open heavy flavor have established a mass dependence of parton energy loss, demonstrated substantial collective flow of heavy quarks, and provided increasingly stringent constraints on heavy-quark diffusion and hadronization. Heavy quark diffusion in the QGP medium offers a microscopic picture to the inner structure of ``perfect liquid".
Forward and backward measurements in proton-nucleus collisions have clarified the importance of cold nuclear matter effects, while charm-baryon production has emerged as a powerful probe of hadronization through coalescence.

Quarkonium measurements have revealed the intricate interplay of cold nuclear matter effects, dissociation, and regeneration. Sequential suppression across the charmonium and bottomonium families provides compelling evidence for medium-induced modification of the heavy-quark potential and offers a unique window into the thermodynamic properties of the QGP.

Although collider operations have ended, the RHIC heavy-flavor program is entering its precision era. Capitalizing on the large datasets and new detector capabilities by the STAR and \sphenix experiments will advance precision measurements of open charm, open bottom, and quarkonia over the coming decade. Looking further ahead, the EIC will extend these studies by providing precision measurements of heavy-flavor production in (polarized) \ep\ and $e$+A collisions, enabling a clean separation of cold nuclear matter effects from hot-medium phenomena. Together, RHIC and the EIC will provide a comprehensive understanding of heavy-quark production, transport, and hadronization across the full landscape of QCD.

%
% Each of the commands below will create an unnumbered section with the appropriate heading.
% Remove any sections that are not relevant for your article.
% All sections except suppdata will be removed if the [anonymous] option is used.
% See iopjournal-guidelines.pdf for more information.
%

%\ack{Sample text inserted for demonstration.}

%\funding{Sample text inserted for demonstration.}
% This section is a list of funder names and grant numbers

%\roles{Sample text inserted for demonstration.}
% List author names and the contributions made to the article, using terms from the NISO Contributor Roles Taxonomy (CRediT) https://credit.niso.org

%\data{Sample text inserted for demonstration.}
% For more information on IOP Publishing's research data policy see: https://publishingsupport.iopscience.iop.org/questions/research-data/

%\suppdata{Sample text inserted for demonstration.}

%\section*{References}
%\biboptions{numbers,sort&compress}
\bibliographystyle{iopart-num} 
\bibliography{references}

\end{document}